\documentclass{jfm}
\usepackage{graphicx}
\usepackage{epstopdf, epsfig}
\usepackage{upgreek} 
\usepackage{bm} 
\usepackage{mathtools} 
\usepackage{mathrsfs}
\usepackage{color}
\usepackage{xcolor}
\usepackage{tabularx}
\usepackage{comment}

\newcommand\putfig[2]{\begin{tabular}[t]{@{}l@{}}#1\\#2\end{tabular}}

\shorttitle{VLSM modulation by inertial particles in open channel flow}
\shortauthor{G. Wang and D. H. Richter}

\title{Two mechanisms of VLSM modulation by inertial particles in open channel flow}
\author{
  G. Wang
 \and 
  D. H. Richter
  \corresp{\email{David.Richter.26@nd.edu}}  
 }

\affiliation{Department of Civil and Environmental Engineering and Earth Sciences, University of Notre Dame, Notre Dame, Indiana 46556, USA}

\begin{document}

\maketitle

\begin{abstract}
Very large-scale motions (VLSMs) and large-scale motions (LSMs) coexist at moderate Reynolds numbers in a very long open channel flow. Direct numerical simulations two-way coupled with inertial particles are analysed using spectral information to investigate the modulation of VLSMs. In the wall-normal direction, particle distributions (mean/preferential concentration) exhibit two distinct behaviors in the inner flow and outer flow, corresponding to two highly anisotropic turbulent structures, LSMs and VLSMs. This results in particle inertia's non-monotonic effects on the VLSMs: low inertia (based on the inner scale) and high inertia (based on the outer scale) both strengthen the VLSMs whereas moderate and very high inertia have little influence. Through conditional tests, low and high inertia particles enhance VLSMs following two distinct routes. Low inertia particles promote VLSMs indirectly through the enhancement of the regeneration cycle (the self-sustaining mechanism of LSMs) in the inner region whereas high inertia particles enhance the VLSM directly through contribution to the Reynolds shear stress at similar temporal scales in the outer region. This understanding also provides more general insight into inner-outer interaction in high Reynolds number, wall-bounded flows. 
\end{abstract}

\begin{keywords}
VLSM, modulation, open channel, DNS, inertial particles
\end{keywords}
\section{Introduction}\label{sec:Introduction}

Very large-scale motions (VLSMs) extending to over $20 h$ (where $h$ is the boundary layer thickness) are found in very high Reynolds number, wall-bounded turbulent flows and are distinct from the well-understood large-scale motions (LSMs) which form canonical streaks and hairpin vortices \citep{hutchins2007evidence,smits2011high,jimenez2011cascades}. These long, meandering features are observed to be energetic, carrying $40-65\%$ of the kinetic energy and $30-50\%$ of the Reynolds shear stress in pipe flow \citep{balakumar2007large} or in turbulent boundary layers \citep{lee2011very}, which is contradictory to the notion of ``inactive" motion proposed by \cite{townsend1980structure}. In environmental flows, these anisotropic structures also have significant influence on the dispersion of pollutants, sand, and other constituents. At the same time, understanding the modulation of turbulence by inertial particles is itself a formidable challenge \citep{balachandar2010turbulent}, and nearly all numerical studies of two-way coupling in particle-laden wall turbulence have been restricted to low Reynolds numbers. It is therefore the aim of this investigation to study the effects of particles on VLSMs, in particular focusing on the question of whether particles act directly or indirectly on these very large motions.\\

\begin{figure}
\centering
\putfig{}{\includegraphics[width=13.5cm,trim={0cm 0cm 0cm 0cm}, clip]{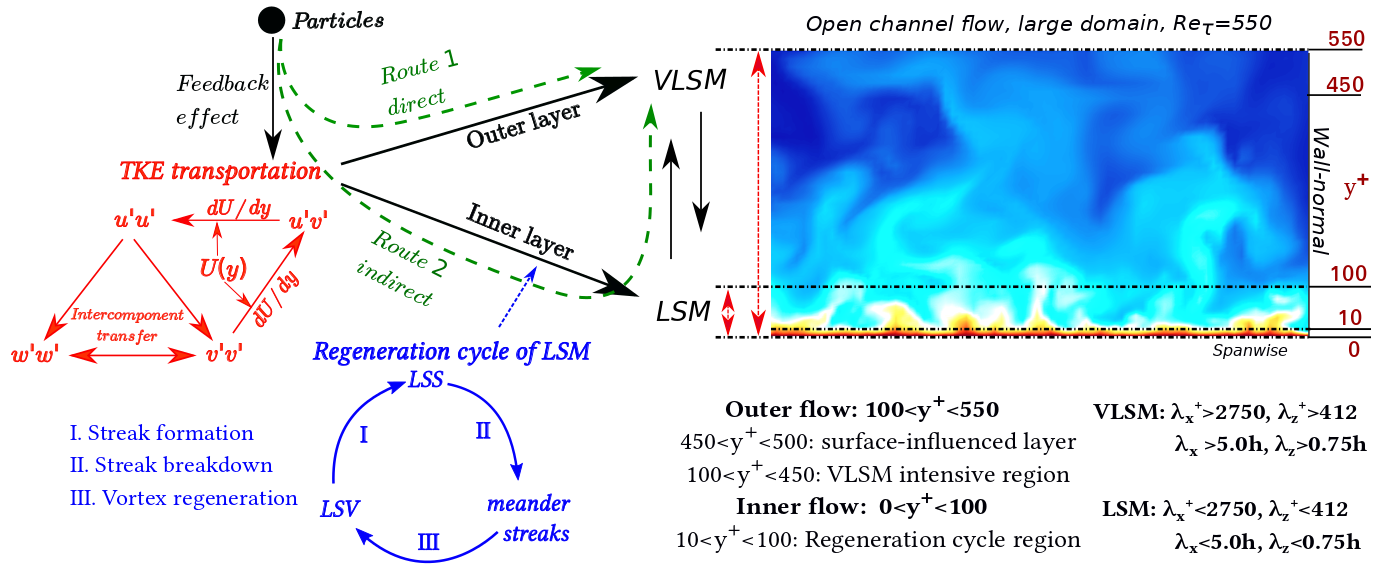}}
  \caption{Schematic of two routes of inertial particle effects on VLSMs through direct impact on TKE transportation or via indirect upscale energy transfer from LSMs. In the right is streamwise velocity contour in the cross-stream plane of $\Rey_\uptau=550$ open channel flow (the flow simulated here) and the flow regions. The scale of VLSMs and LSMs definition used in this study is based on \cite{del2003spectra}. In the bottom-left is the illustration of regeneration cycle of near-wall turbulent structures proposed by \cite{hamilton1995regeneration}.}
\label{fig:sketch_interaction}
\end{figure}
In contrast with with VLSMs, the importance of LSMs on the flow dynamics in the near-wall region has been demonstrated in many different contexts. The LSMs are found to follow a self-sustaining process (i.e. a regeneration cycle) characterized by three key structures shown in the lower left of figure \ref{fig:sketch_interaction}: large-scale streaks (LSSs), large-scale vortices (LSVs), and meandering streaks. Associated with these structures are three regeneration processes: streak formation, streak breakdown, and vortex regeneration \citep{hamilton1995regeneration, waleffe1997self, schoppa2002coherent}. The particle modulation of the regeneration cycle has been investigated by \cite{brandt2014lift}, \cite{wang_abbas_climent_2018} and \cite{wang2019modulation} to explain the non-monotonic effect of the onset of transition with mass loading and particle size. The typical scales of VLSMs, meanwhile, are far larger than LSMs, with their spanwise wavelength $\lambda_z \geq h$ and their streamwise wavelength $\lambda_x \geq 10h$ \citep{Kim1999PoF, guala2006large, adrian2012coherent}. These structures co-exist with LSMs, and the interaction between them is still an open question. It is generally accepted that the regeneration cycle of LSMs does not require the existence of VLSMs \citep{jimenez1999autonomous, hwang2016self}, and \cite{guala2006large} propose that the formation of LSMs and VLSMs results from different mechanisms. \cite{rawat2015self} argue that VLSMs are self-sustained and do not draw energy from LSMs in the buffer layer, however, \citet{Kim1999PoF} and \cite{adrian2012coherent} suggest that VLSMs are not a new type of turbulent structure but merely the consequence of the alignment of coherent LSMs. \cite{toh2005interaction} show numerically that LSMs and VLSMs interact in a co-supporting cycle. Recent work by \cite{lee2019spectral} reveals that very near the wall ($y^+ < 15$) there is a true inverse scale transfer from the dominant LSMs to VLSMs, which appears to be driven by interaction of the streaks with large-scale outer-layer structures. In the current work, we find that the enhancement of VLSMs can be caused by the promotion of LSMs via inertial particles.\\

A wide ranging parameter space including flow type, mass fraction, and particle-to-fluid length and time scale ratios can be expected to complicate the picture of turbulence modulation, resulting in poorly mapped out mechanisms of particle two-way coupling \citep{balachandar2010turbulent}. In isotropic turbulence, \cite{poelma2006particle} reviewed particle influences on the turbulent kinetic energy (TKE) spectrum of the carrier fluid, indicating that low wavenumbers are suppressed, while energy is gained at higher wave numbers. The physical explanations are still not well understood, however. The situation is even more complex for wall-bounded turbulence, especially in the logarithmic layer which contains a multiscale momentum cascade in three-dimensional space \citep{jimenez2011cascades, jimenez2018coherent, marusic2019attached}, while the particle-to-fluid length and time scale ratios also vary as a function of wall-normal height. In addition, the wall-normal TKE transport is also modulated by inertial particles, for example, \cite{zhao2013interphasial} found that inertial particles transported within streaky motions act as a carrier transferring TKE from the core region of the channel to the fluid close to the wall. There has been substantial progress in understanding inertial particle dynamics in the inner layer (i.e. related to LSMs), for instance the phenomena of particle clustering and segregation \citep{pan1995numerical, marchioli2002mechanisms, sardina2012wall}, drag reduction \citep{dritselis2008numerical}, high particle loading and cluster dynamics \citep{capecelatro2018transition}, particle inducing upscale energy transfer and transportation \citep{richter2015turbulence}, and regeneration cycle modulation \citep{wang2019modulation}. However to the best of our knowledge, very little attention has been paid to particle clustering and modulation of VLSMs in the outer layer. \\

More generally, it is inherently difficult to describe turbulence modulation by particles. The intensity of TKE is often used to indicate turbulence modulation (e.g. \cite{Pan1996PoF}, \cite{crowe2000models} and \cite{tanaka2008classification}), however, this can lead to somewhat contradictory descriptions. For instance, recent observations have shown that even though TKE is nearly unchanged, the onset of laminar-to-turbulent transition can be significantly advanced by particles with a low Stokes number ($St^+=\mathcal{O}(1)$; based on viscous time scale); see for example \cite{klinkenberg2013numerical}, \cite{wang2019modulation}, \cite{saffman1962stability}, and \cite{michael1964stability}. In fact, not all scales of turbulence are enhanced during turbulence augmentation, so using the bulk TKE might misrepresent the modulation at certain length and time scales. This can be observed with spectral analysis, which is a natural means to study particles modulation of turbulence. For example \cite{elghobashi1993two} found there is a possible so-called reverse cascade which tends to build up energy in large scale structures in homogeneous turbulence, and \cite{richter2015turbulence} demonstrates that this upscale influence is a strong function of particle inertia. As a result, particles can influence turbulence scales far removed from their own response time scale in wall-bounded turbulence. \\

In this study, as indicated in figure \ref{fig:sketch_interaction}, particles can directly impact TKE transport through momentum coupling \citep{elghobashi1993two}, thereby modulating specific scales of turbulent structures directly, i.e. VLSMs in the outer layer (through route 1 in figure \ref{fig:sketch_interaction}) or LSMs in the inner layer. As discussed above, particle modulation of LSMs in the inner layer also has the possible effect that particle feedback on LSMs near the wall can have upscale, indirect influences on VLSMs via nonlinear energy transfer (through route 2 in figure \ref{fig:sketch_interaction}); see for example \cite{toh2005interaction} and \cite{lee2019spectral}. To better understand particle modulation of LSMs in the inner layer, \cite{wang2019modulation} further investigated small particles and their ability to enhance the LSM regeneration cycle (depending non-monotonically on particle inertia, see also \cite{saffman1962stability}), with the assumption that this was a route through which particles could modify even larger scales in high Reynolds number flow. Therefore as a follow-up, and since to date computational costs have precluded particle-laden direct numerical simulations at sufficiently high Reynolds number, we for the first time examine the effects of a wide range of particle inertia on VLSMs in open channel flow at Reynolds numbers of $\Rey _\uptau=550$ and $\Rey _\uptau=950$.\\

\section{Numerical parameters}\label{sec:Parameters}
\begin{table}
\centering
\begin{tabular}{cccccccccccc}
\multicolumn{12}{l}{Type 1 ($\Rey_{\uptau}=550$)}                                                                                                       \\
\multicolumn{12}{c}{$N_x \times N_y \times N_z = 1024 \times 128 \times 512$}                                                                           \\
\multicolumn{12}{c}{$L_x \times L_y \times L_z = 6\pi \times 1 \times 2\pi$}                                                                            \\
\multicolumn{12}{c}{$L^+_x \times L^+_y \times L^+_z = 10367 \times 550 \times 3456$}                                                                   \\
\multicolumn{12}{c}{$\Delta x^+ \times \Delta y^+ (wall,~surface) \times \Delta z^+=10.1 \times (1,~7.2)  \times 6.75$}                                 \\
\multicolumn{12}{l}{Type 2 ($\Rey_{\uptau}=550$)}                                                                                                       \\
\multicolumn{12}{c}{$N_x \times N_y \times N_z = 2048 \times 128 \times 512$}                                                                           \\
\multicolumn{12}{c}{$L_x \times L_y \times L_z = 12\pi \times 1 \times 2\pi$}                                                                           \\
\multicolumn{12}{c}{$L^+_x \times L^+_y \times L^+_z = 20734 \times 550 \times 3456$}                                                                   \\
\multicolumn{12}{c}{$\Delta x^+ \times \Delta y^+ (wall,~surface) \times \Delta z^+=10.1 \times (1,~7.2)  \times 6.75$}                                 \\
\multicolumn{12}{l}{Type 3 ($\Rey_{\uptau}=950$)}                                                                                                       \\
\multicolumn{12}{c}{$N_x \times N_y \times N_z = 1024 \times 256 \times 512$}                                                                           \\
\multicolumn{12}{c}{$L_x \times L_y \times L_z = 10.8 \times 1 \times\pi$}                                                                              \\
\multicolumn{12}{c}{$L^+_x \times L^+_y \times L^+_z = 10260 \times 950 \times 2984$}                                                                   \\
\multicolumn{12}{c}{$\Delta x^+ \times \Delta y^+ (wall,~surface) \times \Delta z^+=10 \times (1,~6.4)  \times 5.8$}                                    \\
\multicolumn{12}{c}{ }                                    \\
Type & $Num.$ & $\overline{\Phi_m}$ & $\rho_p / \rho_f$ & $\overline{\Phi_v}$           & $N_p$           &  & $\uptau_p$ & $St^+$ & $St_K$         & $St_{out}$ & $u_\uptau T/h$ \\
     &        &          &                 & $(\times 10^{-4})$ & $(\times 10^6)$ &  &            &        & $(inner,outer)$ &          &                \\
1    & case1  & \multicolumn{9}{l}{single-phase flow}                                                                                  & 32             \\
     & case2  & 0.024    & 16              & 15                 & 12.6            &  & 0.51       & 2.42   & $0.587,~0.211$  & 0.08     & 22             \\
     & case3  & 0.14     & 160             & 8.75               & 7.33            &  & 5.1        & 24.2   & $5.87,~2.11$   & 0.8      & 22             \\
     & case4  & 0.14     & 400             & 3.5                & 2.93            &  & 12.7       & 60.5   & $14.7,~5.26$   & 2.0      & 22             \\
     & case5  & 0.14     & 1200            & 1.17               & 0.98            &  & 38.2       & 182    & $44.0,~15.8$   & 6.0      & 22             \\
     & case6  & 0.14     & 6000            & 2.33               & 0.195           &  & 191        & 908    & $220,~79$     & 30       & 22             \\
2    & case7  & \multicolumn{9}{l}{single-phase flow}                                                                                       & 20             \\
3    & case8  & \multicolumn{9}{l}{single-phase flow}                                                                                       & 18             \\
     & case9  & 0.14     & 1600            & 0.875              & 1.68            &  & 12.7       & 180.5    & $45.1,~12.8$  & 4.1      & 18             \\
     & case10 & 0.14     & 3200            & 0.438              & 0.84            &  & 25.5       & 361    & $90.3,~25.5$    & 8.2      & 18            
\end{tabular}

  \caption{Parameters of numerical simulations. The friction Reynolds number is $\Rey_\tau\equiv u_{\uptau}h/\nu$ where $h$ is the depth of the open channel and the particle relaxation time is $\uptau _p \equiv \rho_p d^2/(18\rho_f \nu)$ where $d$ is the particle diameter. The ratio $d_p/\eta_K$ is maintained at a value of approximately $0.42$. The particle Reynolds number remains $\mathcal{O}(1)$ or lower. $\overline{\Phi_m}$ is the particle mass concentration and $N_p$ is the total particle number. The superscript ``+" is the dimensionless number based on viscous scale, where $\delta_\nu$, $u_{\uptau}$ and $\nu/ u^2_{\uptau}$ correspond to the viscous length scale, velocity scale, and time scale, respectively. $St_K$ represents the dimensionless particle time scale based on averaged Kolmogorov time scale in the inner layer $y^+<100$ and outer layer $y^+>100$, corresponding to LSMs and VLSMs strong region as shown in figure \ref{fig:kzEuu_2d_550_950}. $St_{out}$ represents the dimensionless particle time scale based on the outer flow time scale $h/U_{bulk}$, where $U_{bulk}$ is the bulk velocity of the channel.}
  \label{tab:Table_1} 
\end{table}

\begin{figure}
\centering
\putfig{}{\includegraphics[width=13.5cm,trim={0cm 0cm 0cm 0cm}, clip]{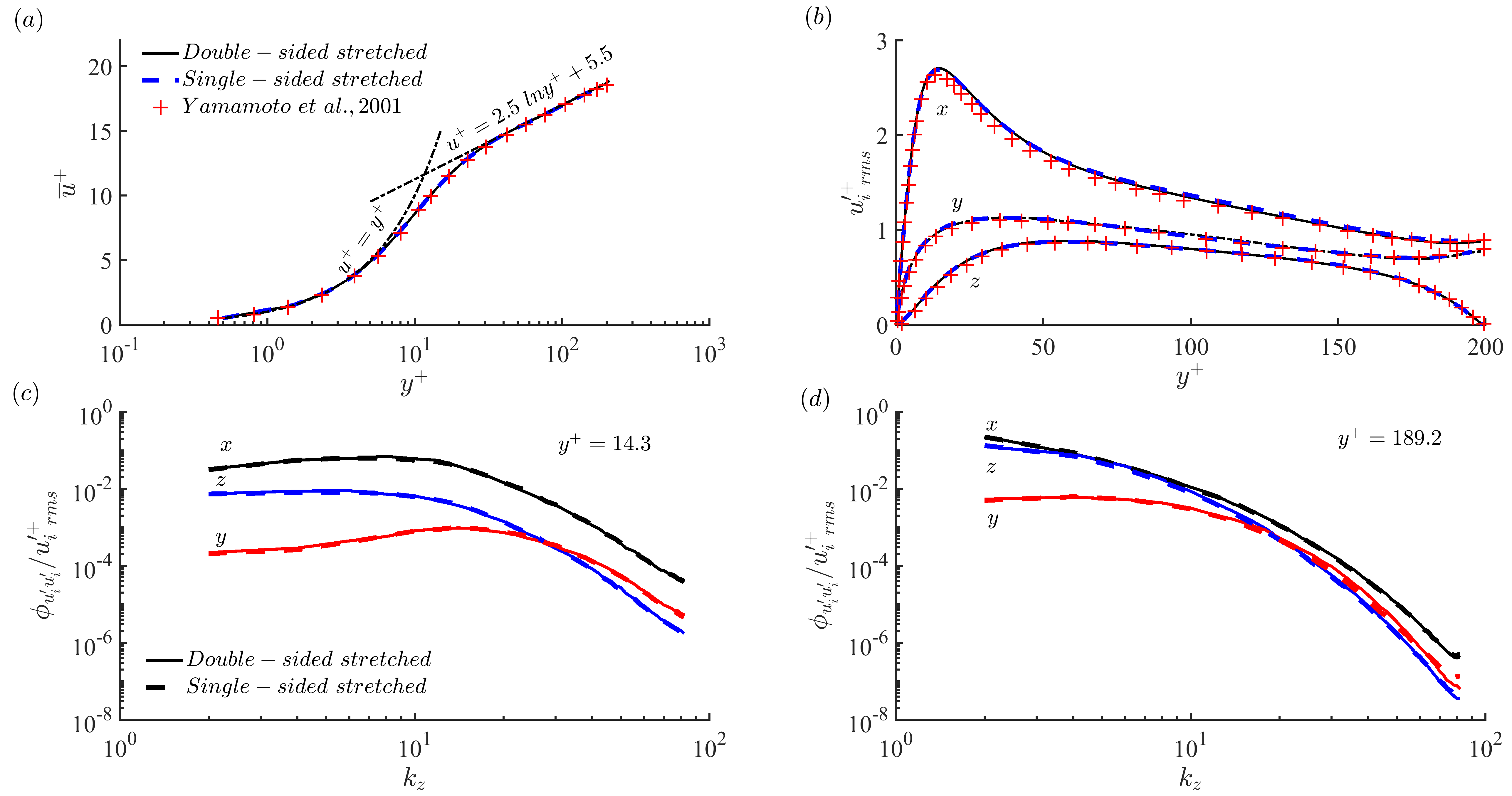}}
  \caption{$(a)$ Mean streamwise velocity profile. $(b)$ Profiles of oot-mean-square of velocity fluctuations. $(c,d)$ streamwise TKE spectra as function of wavenumber in spanwise direction at two wall-normal heights: $(c)$ close to the wall $y^+=14.3$ and $(d)$ near free-surface region $y^+=189.2$. All figures are normalized by viscous scales.}
\label{fig:mesh_Yamamoto}
\end{figure}

Direct numerical simulations of the Eulerian flow are performed for an incompressible Newtonian fluid using the same numerical implementation as \citet{richter2014modification} and \citet{richter2015turbulence}. A pseudospectral method is employed in the periodic directions (streamwise $x$ and spanwise $z$), and second-order finite differences are used for spatial discretization in the wall-normal ($y$) direction. We simulate pressure-driven open channel flow because it is characterized by features of both closed channel flow and boundary layer \citep{nezu1993turbulence,cameron2017very}, while also exhibiting the signatures of VLSMs at a more computationally accessible domain size and Reynolds number. A no-slip condition is imposed on the bottom wall and a shear-free condition is imposed on the upper surface, and such boundary conditions have been proven capable of capturing many of the phenomena (e.g. VLSMs) seen in experiments with shear-free upper boundaries; see \cite{pan1995numerical,Pan1996PoF, adrian2012coherent}. We remark here that the inertial particles do not collect at the free surface in this study, which is also observed experimentally by \cite{sumer1978particle} and numerically by \cite{Pan1996PoF}. The solution is advanced in time by a third-order Runge-Kutta scheme. A single-sided stretched grid (fine grid close to the wall, coarse grid close to the free surface) is used in this study. Comparisons with a double-sided stretched grid (fine grid close to the wall and the free surface) and the simulations of \cite{yamamoto2001turbulence} at $\Rey_\uptau=200$ produce nearly identical mean velocity profiles, shown in figure \ref{fig:mesh_Yamamoto}(a), and turbulent intensity profiles, shown in figure \ref{fig:mesh_Yamamoto}(b). In addition, the one-dimensional $u-$spectra $\phi_{u'u'}(k_z)=\langle{\hat{u'}(k_z)\hat{u'}^*(k_z)}\rangle$ for the single-sided stretched and double-sided stretched grids is shown in figure \ref{fig:mesh_Yamamoto} (c-d). Both close to the wall ($y^+=14.3$; figure \ref{fig:mesh_Yamamoto}(c)) and near free-surface region ($y^+=189.2$; figure \ref{fig:mesh_Yamamoto}(d)), single-sided and double-sided stretched grids agree with each other. \\

Particle trajectories and suspension flow dynamics are based on the Lagrangian point-particle approximation where the particle-to-fluid density ratio $\rho_p / \rho_f \gg 1$ and the particle size is smaller than the smallest viscous dissipation scales of the turbulence. Only the Stokes drag force and two-way coupling have been incorporated since we restrict our study to low volume concentration $\overline{\Phi_{V}}=\mathcal{O}(10^{-4})$ \cite[see][]{balachandar2010turbulent}. Gravitational settling is not considered in order to highlight the effect of the particle response time. Particles experience a purely elastic collision with the lower wall and upper rigid free surface. Two-way coupling is implemented via a particle-in-cell scheme, and has been validated against \cite{zhao2013interphasial} and \cite{capecelatro2013euler} in turbulent channel flow. Grid convergence of both the flow and of the two-way coupling scheme have been verified as well \citep{gualtieri2013clustering}.\\

 Particle modulation of turbulence is often characterized by the relative time scales between particles and local turbulent structures. The multiple turbulent structures spanning a wide spatial and temporal range (e.g. LSMs and VLSMs) result in a wide parameter space of the particle inertia to be investigated. As shown in table \ref{tab:Table_1}, we choose $St^+$ in the range of $2.42 - 908$ based on the inner viscous time scale, which corresponds to $St_{out}$ ranging from $0.08 - 30$, where $St_{out}$ is based on the outer bulk flow time scale $h/U_{bulk}$. This also corresponds to $St_K$ in the range $0.587 - 220$ based on the average Kolmogorov scale in the inner layer and $0.211 - 79$ based on the average Kolmogorov scale in the outer layer. In single-phase channel flow, \cite{del2003spectra} use $L_x \times L_z= 8 \pi h \times 4 \pi h$ at $\Rey_\uptau=550$ and \cite{abe2004very} choose $L_x \times L_z= 12.8 h \times 6.4 h$ at $\Rey_\uptau=640$ to study VLSMs. In current particle-laden flow, the domain size $L_x \times L_z= 6 \pi h \times 2 \pi h$ is used, slightly shorter than \cite{del2003spectra} whereas larger than \cite{abe2004very}. With this domain size, we observe the appearance of a bimodal energy spectra in the spanwise direction and compare well with \cite{del2003spectra} (see figure \ref{fig:kzEuu_2d_550_950}). In single-phase flow it is well-known that VLSMs are very long in the streamwise direction and fully capturing their extent is computationally expensive \citep{lozano2014effect}. Therefore as a test, $case7$ doubles the streamwise extent for single-phase flow in order to check any effects of streamwise confinement on VLSMs by comparing to $case1$ (streamwise velocity spectrum in spanwise direction is shown later in figure \ref{fig:kzEuu_2d_550_950}(a); negligible differences were observed). $Cases~1-6$ are then designed to investigate the effects of particle inertia by systematically increasing the particle Stokes number. In order to further examine particle direct modulation of VLSMs, $cases~8-10$ are performed at a higher $\Rey_\uptau=950$ for single-phase and particle-laden flow --- these ultimately yield identical conclusions. \\

\section{Results}\label{sec:Results}

\subsection{Particle distribution in two distinct layers}\label{subsec:particle_distribution}

\begin{figure}
\centering
\putfig{}{\includegraphics[width=13.5cm,trim={0cm 0cm 0cm 0cm}, clip]{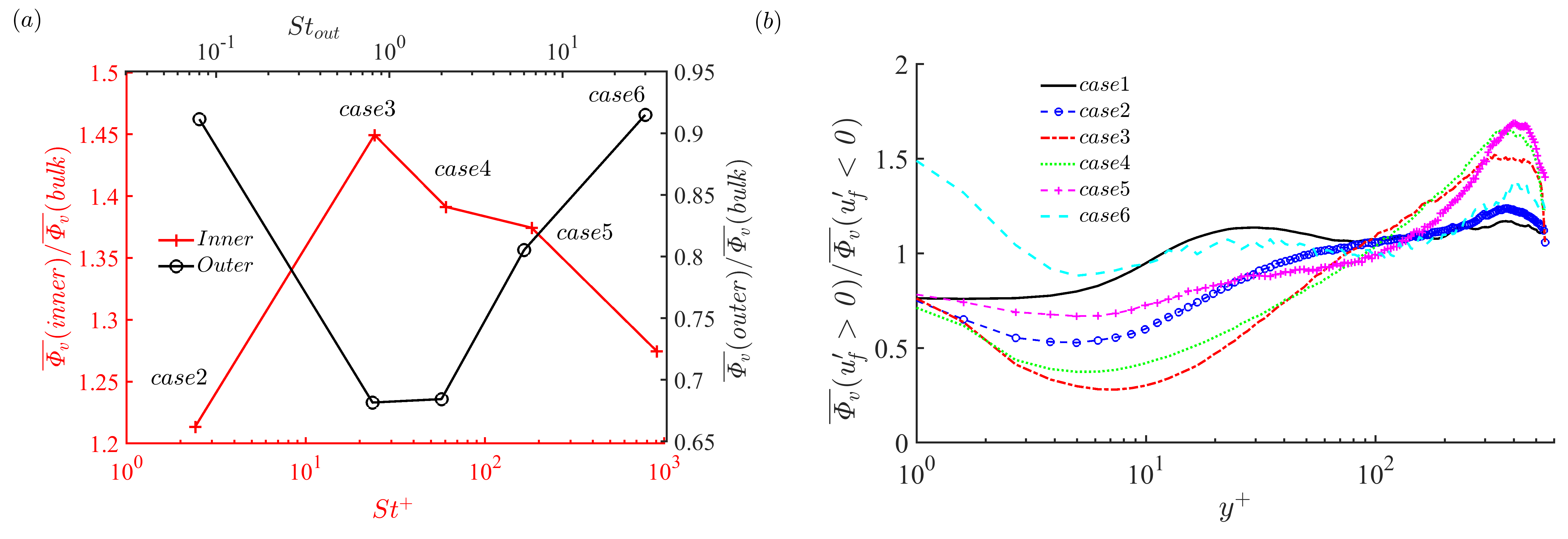}}
  \caption{$(a)$ Mean particle volume concentration in the inner layer (as a function of $St^+$) and outer layer (as a function of $St_{out}$), scaled by the bulk value. $(b)$ The ratio between particle concentrations with $u'_f>0$ and $u'_f<0$. For single-phase flow ($case1$), Eulerian grid points with $u'>0$ or $u'<0$ are plotted.}
\label{fig:Concentration_high_low_streak}
\end{figure}

Mean particle volume concentrations in the inner layer and outer layer are shown in figure \ref{fig:Concentration_high_low_streak}(a), exhibiting a non-monotonic behaviour with Stokes number but with an opposite trend. In the inner layer it is maximized for $case3$, which corresponds to $St^+=24.2$. At the same time, $case3$ also exhibits the minimum in the outer layer, where the more relevant Stokes number is $St_{out} = 0.8$. Stokes numbers lower or higher than $case3$ result in fewer particles in the inner layer whereas more particles in the outer layer. This often-observed behavior is due to turbophoresis \citep{reeks1983transport}, which induces a net particle flux towards the wall resulting in higher particle volume concentration in the inner layer than in the outer layer. 

It is also commonly accepted that inertial particles preferentially accumulate in low-speed streaks \citep{Pan1996PoF, marchioli2002mechanisms}, which is also observed in this study. Particle numbers in `upwelling'  and `downwelling' regions can be straightforwardly counted by testing whether $u'_f<0$ or $u'_f>0$, where $u'_f$ is the fluid fluctuating velocity seen by the particle. Then, the ratio of the number of particles in `upwelling' and `downwelling' regions can be used to compare across different wall-normal locations and Stokes numbers. This ratio, cast in terms of the effective volume concentration corresponding to these particle counts, is shown in figure \ref{fig:Concentration_high_low_streak}(b). Here, there is a clear non-monotonic trend with the Stokes number and with $y$. In the inner layer, there are more particles in the `upwelling' fluid motions than in the `downwelling' fluid motions, which is opposite compared to the outer layer. The lowest ratio appears for $case3$ at $St^+=24.2$ in the inner layer, while the highest ratio appears for $case5$ at $St_{out}=6.0$ in the outer layer. Thus in the inner layer, relatively low-inertia particles collect in the low-speed streaks, while in the outer layer, higher-inertia particles collect in the high-speed regions. The relevant Stokes numbers are different for each, since the respective fluid timescales are different.\\

\begin{figure}
\centering
\putfig{}{\includegraphics[width=13.5cm,trim={0cm 0cm 0cm 0cm}, clip]{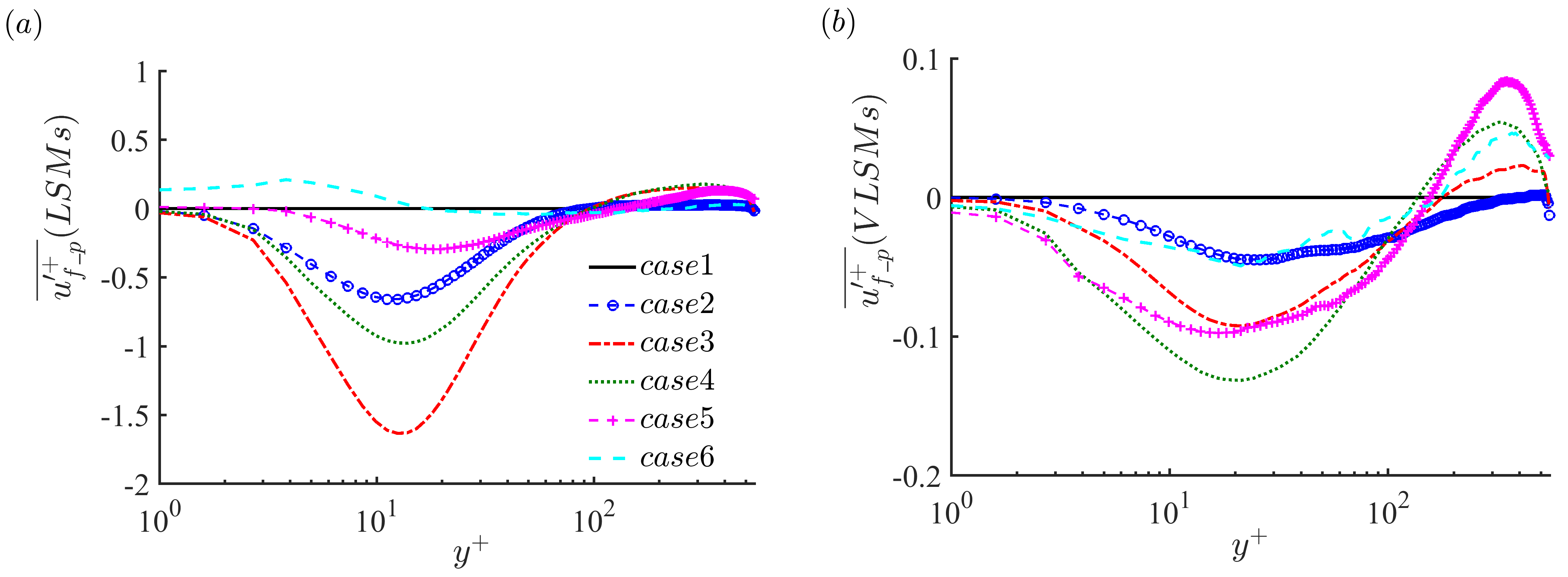}}
  \caption{Mean fluid streamwise velocity fluctuation at particle locations defined in equation \ref{eq:Fluid_p_VLSM_LSM}, $(a)$ in LSMs ($\lambda_x < 5h$, $\lambda_z < 0.75h$) and $(b)$ in VLSMs ($\lambda_x > 5h$, $\lambda_z > 0.75h$).}
\label{fig:Concentration_VLSM}
\end{figure}

The relationship between particle distributions and the different turbulent structures (i.e. VLSMs and LSMs) is still poorly understood, however. To explore this, we introduce a mean fluid streamwise velocity fluctuation at the particle positions, but filtered according to the wavelengths associated with LSMs and VLSMs: 
\begin{equation}\label{eq:Fluid_p_VLSM_LSM}
\overline{u'_{f\_p}}(VLSMs) \equiv \frac{\overline{\Phi_v u'_{\lambda_x > 5h, \lambda_z > 0.75h}}}{\overline{\Phi_v}};~~
\overline{u'_{f\_p}}(LSMs)  \equiv \frac{\overline{\Phi_v u'_{\lambda_x < 5h, \lambda_z < 0.75h}}}{\overline{\Phi_v}}.
\end{equation}
Here the interpolated velocity $u_{f}$ is projected onto the Eulerian grid, allowing for the $u'_{f}$ field  to be transferred to Fourier space ($\hat{u}'_f$). The goal is to artificially filter out targeted turbulent structures (e.g. removing wavelengths with $\lambda_x < 5h$, $\lambda_z < 0.75h$ to isolate VLSMs) in order to obtain $u'_f$ contributed by specific turbulent structures. For single-phase flow, $\Phi_v$ is set to one at all Eulerian grid points and the same procedure is followed. These quantities are shown in figure \ref{fig:Concentration_VLSM}. In the inner layer, $\overline{u'_{f\_p}}$ is negative in both VLSMs and LSMs, indicating that particles are more likely to reside in the large and very large scale low-speed streaks compared to the high-speed streaks. The minimum values of $\overline{u'_{f\_p}}$ for LSMs ($case3$ with $St^+=24.2$ in figure \ref{fig:Concentration_VLSM}(a)) and VLSMs ($case4$ with $St^+=60.5$ in figure \ref{fig:Concentration_VLSM}(b)) appear at different Stokes numbers, again since the flow timescales associated with LSMs and VLSMs are different. 

\begin{figure}
\centering
\putfig{}{\includegraphics[width=13.5cm,trim={0cm 0cm 0cm 0cm}, clip]{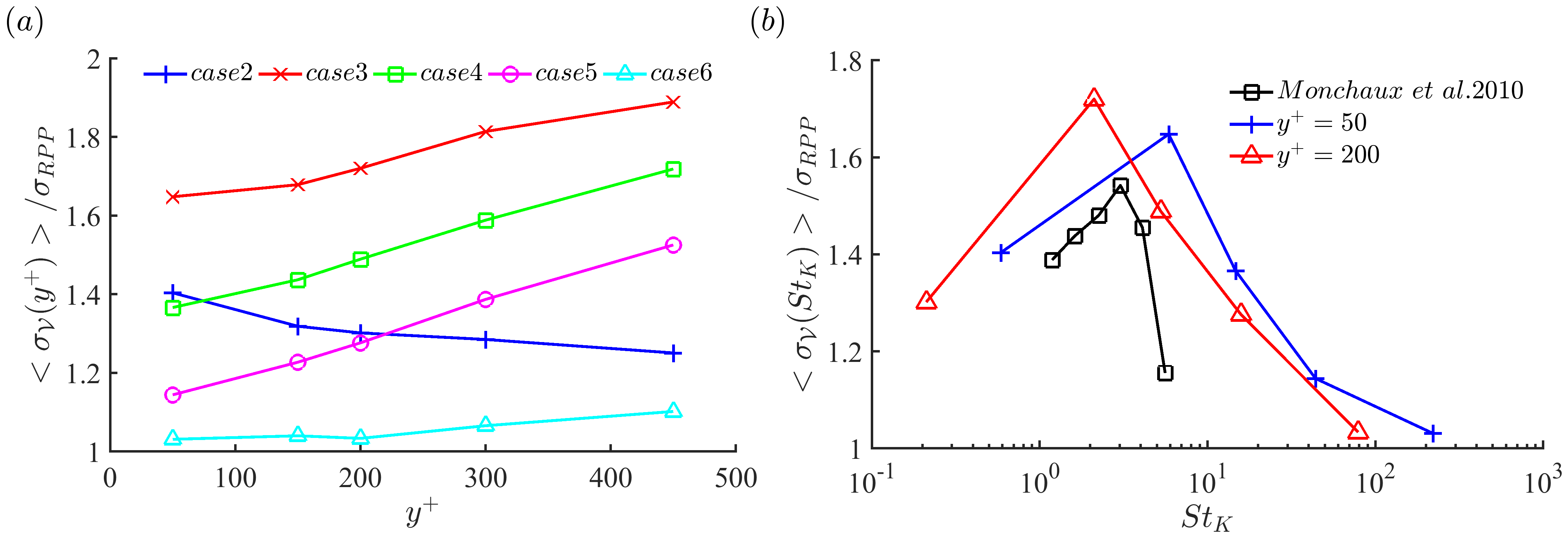}}
  \caption{Standard deviation of the normalized Vorono\"i area $\sigma_{\mathcal{V}}$, normalized by that of a random Poisson process, $\sigma_{RPP}$. $(a)$ as a function of height in wall-normal direction of five Stokes numbers and $(b)$ as a function of $St_K$ at two wall-normal locations ($y^+=50$ and $y^+=200$ are plotted based on $St_{K}$ of the inner layer and outer layer as in table \ref{tab:Table_1}). Experimental results of an isotropic turbulence from \citet{monchaux2010preferential} is shown.}
\label{fig:veronoi_variance}
\end{figure}

In the outer layer, $\overline{u'_{f\_p}}$ is positive for both VLSMs and LSMs, indicating that particles tend to reside in the `downwelling' regions at the scales of both the LSMs and the VLSMs. The maximum value of $\overline{u'_{f\_p}}$ for VLSMs (figure \ref{fig:Concentration_VLSM}(b)) appears for $case5$ with $St_{out}=6.0$. Comparing the inner layer with the outer layer, $\overline{u'_{f\_p}}$ for LSMs is considerably stronger in the inner layer than in the outer layer as shown in figure \ref{fig:Concentration_VLSM}(a). Two possible reasons may explain this. One is that the intensity of LSMs is much weaker in the outer layer than in the inner layer (will be shown later in figure \ref{fig:Euu_VLSM_LSM_St}(a)); the other is that particle preferential concentration is not as strongly correlated with streaky motions in the outer layer as compared to the inner layer (will be shown in figure \ref{fig:veronoi_variance}(a)). However for $\overline{u'_{f\_p}}$ in VLSMs (figure \ref{fig:Concentration_VLSM}(b)), the magnitude is comparable between the inner layer and the outer layer, albeit with opposite preferred signs.\\

\begin{figure}
\centering
\putfig{}{\includegraphics[width=12cm,trim={0cm 0cm 0cm 0cm}, clip]{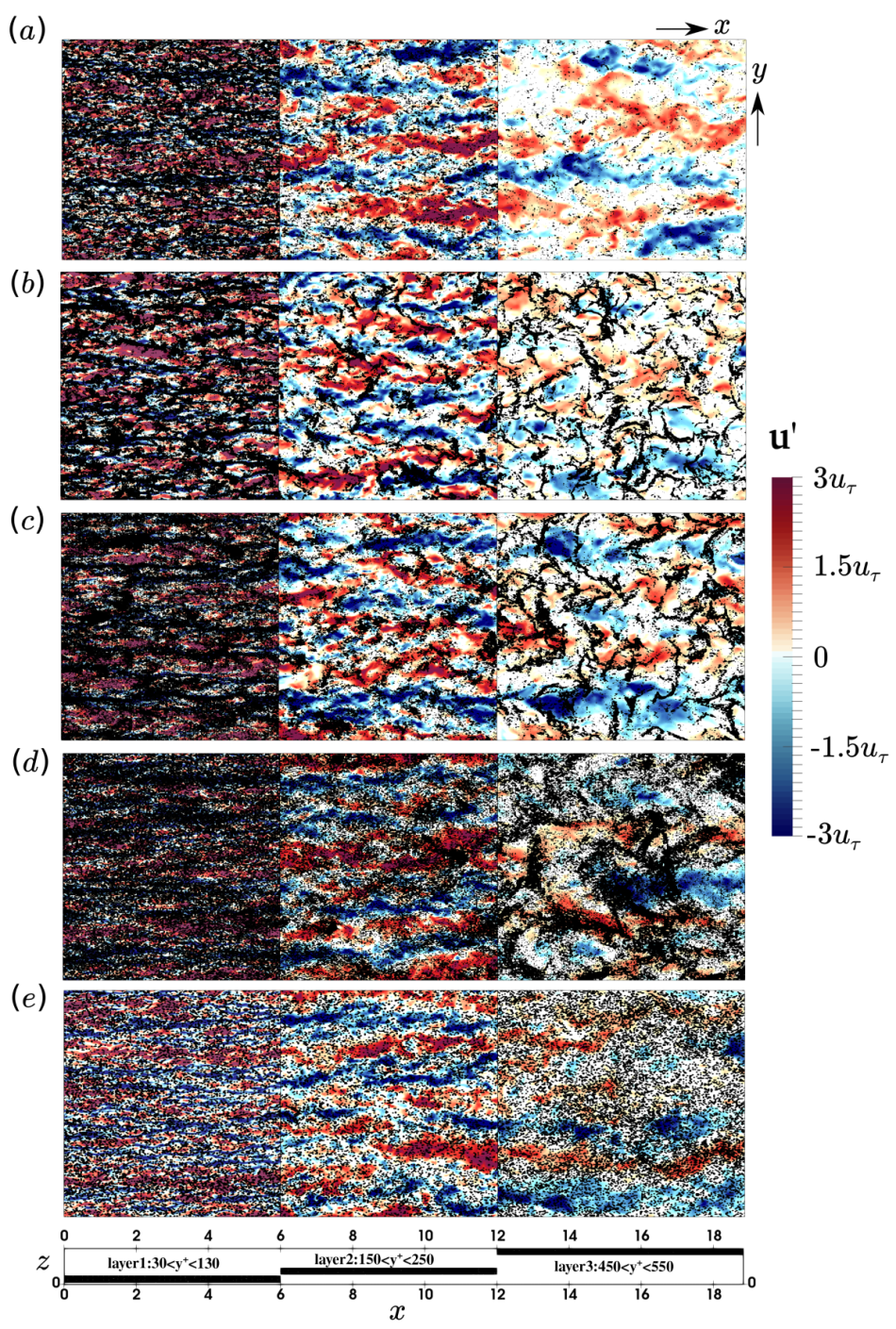}}
  \caption{$(a)$ Instantaneous streamwise velocity fluctuation $u'$ (color contours) in three $x-z$ planes (three panels from left to right: $y^+=50,~150,~450$) and isosurface of particle concentration ($2.5$ times bulk $\overline{\Phi_{v}}$) in three slabs (layer $1$: $30<y^+<130$, layer $2$: $150<y^+<250$, layer $3$: $450<y^+<550$). $(a-e)$ refer to cases $2-6$. A movie is online.}
\label{fig:Concentration_contour}
\end{figure}

In terms of particle clustering behaviour, it is well-established that for wall-bounded turbulent flow in the inner layer, low-inertia particles ($St^+<\mathcal{O}(0.1)$) tend to distribute homogeneously in wall-normal planes \citep{Pan1996PoF}, while intermediate Stokes numbers ($St^+=\mathcal{O}(10)$) exhibit particle clustering in near-wall streaks \citep{marchioli2002mechanisms,sardina2012wall,richter2015turbulence, wang2019modulation} and high-inertia particles ($St^+>\mathcal{O}(100)$) behave with ballistic trajectories (thus eliminating much of the clustering). This qualitative transition with $St^{+}$ is observed within the inner region of the simulated open channel flow. Figures \ref{fig:Concentration_contour}(a-e) present isosurfaces of particle concentration ($2.5$ times the bulk $\overline{\Phi_v}$) for $cases~2-6$ in the inner and outer regions for increasing Stokes numbers. The advantage of showing concentration isosurfaces as opposed to individual Lagrangian points is that this method better visualizes the high-concentration particle clusters. The three panels across the horizontal represent three slabs at progressively increasing wall-normal distances (at the same snapshot in time): layer $1$: $30<y^+<130$, layer $2$: $150<y^+<250$, layer $3$: $450<y^+<550$.

Here we find particles accumulating in the inner-flow low-speed streaks at intermediate $St^+=24.2-60.5$ ($cases~3-4$ in layer $1$ of figures \ref{fig:Concentration_contour}(b-c)); this is similar to many other studies as noted above. At the same time, a new type of organized structure in the outer flow region is observed. These are especially clear at higher Stokes numbers, e.g. $St^+ =182$ ($St_{out} =6$) for $case5$ shown in layers $2-3$ of figure \ref{fig:Concentration_contour}(d). However, with a very high particle inertia, $St^+=908$ ($St_{out} =30$), particles behave ballistically in the outer flow region as shown in layers $2-3$ of figure \ref{fig:Concentration_contour}(e), tending to distribute more homogeneously. These two distinct, non-monotonic particle accumulation behaviours in the inner and outer layers peak at different Stokes numbers ($St^+ =\mathcal{O}(10)$ in the inner layer and $St_{out}=6$ in the outer layer) and have a strong influence on the non-monotonic modulation of the VLSMs in the outer region via the two routes indicated in figure \ref{fig:sketch_interaction}. This will be discussed further in section \ref{subsec:u-spectra}.\\


In order to further quantify the particle clustering behaviour, we employ a Vorono\"i diagram analysis, which compares the distribution of the tessellation areas in the particle-laden cases with the expected Poisson distribution if the particles were randomly distributed (see for example \cite{monchaux2012analyzing}). A maximum clustering effect is typically observed for $St_K$ around unity in isotropic turbulence \citep{monchaux2010preferential, baker2017coherent} and buoyancy-driven wall-bounded turbulence \citep{park2018rayleigh}. Figures \ref{fig:veronoi_variance}(a,b) show the standard deviation ($\sigma_\mathcal{V}$) of the distribution of the normalized Vorono\"i area $\mathcal{V}=A/\overline{A}$, where the inverse of the average Vorono\"i area $\overline{A}$ indicates the mean particle concentration. $\sigma_\mathcal{V}$ is scaled by the standard deviation of a random Poisson process (RPP; $\sigma_{RPP}=0.52$). The ratio $\sigma_\mathcal{V}/\sigma_{RPP}$ exceeding unity indicates that particles are accumulating in clusters as compared to truly randomly distributed particles. 

Figure \ref{fig:veronoi_variance}(a) shows ratio $\sigma_\mathcal{V}/\sigma_{RPP}$ for multiple heights across all Stokes numbers, while figure \ref{fig:veronoi_variance}(b) shows the ratio as a function of $St_{K}$ for two different representative heights ($y^+=50$ and $y^+=200$ are plotted based on $St_{K}$ of the inner layer and outer layer, respectively, as provided in table \ref{tab:Table_1}). In the inner layer ($y^+=50$), from $case2$ to $case6$ ($St^+=2.42-908$ or $St_K=0.58-220$), the clustering effect experiences a non-monotonic evolution as a function of Stokes number. The largest value appears at $St^+=24.2$ and it gradually decreases to unity with a higher Stokes number. This is similar with the investigation of \cite{wang2019modulation} that particles with $St^+=29.5$ preferentially accumulate in the streaks whereas higher inertial particles tend to spread throughout the inner layer. With increasing wall-normal distance, the ratio $\sigma_\mathcal{V}/\sigma_{RPP}$ decreases at very low Stokes number ($case2$) whereas it increases in higher Stokes numbers ($case~3-6$). At still higher Stokes numbers ($case7$), the ratio $\sigma_\mathcal{V}/\sigma_{RPP}$ again approaches unity. These very heavy particles cannot follow the streamlines, resulting in a nearly particle random distribution. In the outer layer ($y^+=200$), the largest value appears at $St_K=2.11$ ($St_{out}=0.8$), similar in magnitude with previous investigations of clustering in isotropic turbulence; peaks were found for $St_K=3.0$ and $St_K=1.0$ by \cite{monchaux2010preferential} and \cite{baker2017coherent}, respectively.

\subsection{Particle modulation of TKE in LSMs and VLSMs}\label{subsec:u-spectra}

\begin{figure}
\centering
\putfig{}{\includegraphics[width=14.0cm,trim={2cm 0cm 1cm 0}, clip]{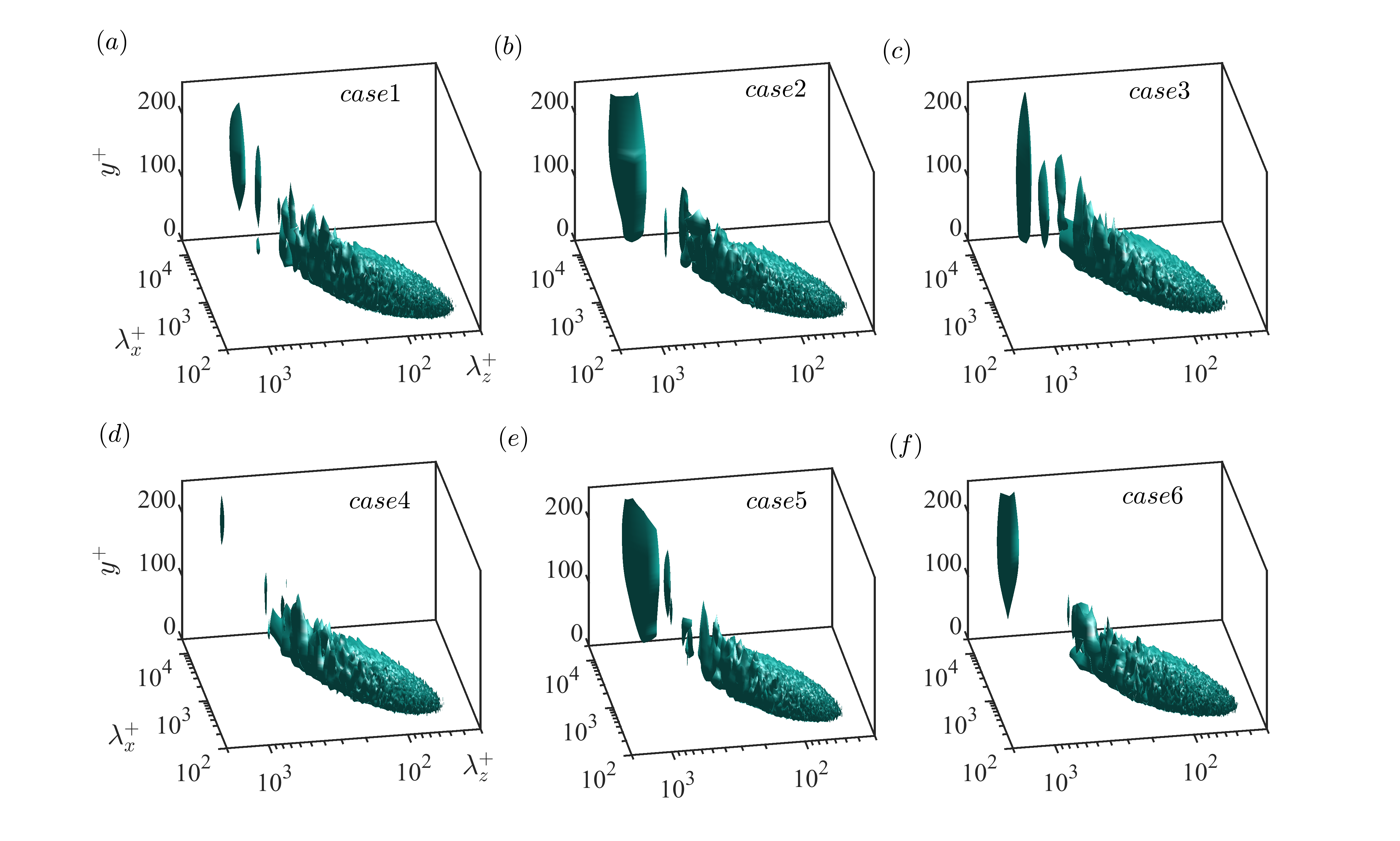}}
  \caption{Premultiplied two-dimensional energy spectrum $k_xk_z\Phi_{u'u'}/u_\uptau ^2$ as a function of $\lambda_x$ and $\lambda_z$ and the wall-normal direction $y$. Isosurface of $0.1$ times the maximum value of the single-phase flow is illustrated. $(a-f)$ refer to cases $1-6$.}
\label{fig:Euu_3d_550}
\end{figure}

The premultiplied, two-dimensional energy spectrum of streamwise velocity, $k_xk_z\Phi_{u'u'}$ where $\Phi_{u'u'}=\langle{\hat{u'}(k_x,k_z,y)\hat{u'}^*(k_x,k_z,y)} \rangle$, is shown in figure \ref{fig:Euu_3d_550} for $\Rey_{\uptau}=550$ as a function of wall-normal distance ($\hat{u'}$ is the Fourier coefficient of $u'$). The figure exhibits a ``boot-shaped" structure, particularly well-defined for $case2$ (figure \ref{fig:Euu_3d_550}(b)) and $case5$ (figure \ref{fig:Euu_3d_550}(e)). The ``forefoot" corresponds to the LSMs in the near-wall region whereas the ``bootleg" corresponds to the VLSMs. The signature of VLSMs indeed appears at the upper-left corner (long and wide wavelengths in the streamwise and spanwise directions) for single-phase flow in figure \ref{fig:Euu_3d_550}(a). This VLSM signature is nearly unchanged for $case3$ and $case6$ whereas it is slightly weakened in $case4$. It is clear, however, that for low Stokes number ($St^+=2.42$ of $case2$ in figure \ref{fig:Euu_3d_550}(b)) and high Stokes number ($St_{out}=6$ of $case5$ in figure \ref{fig:Euu_3d_550}(e)), energy contained by the VLSMs is enhanced by the presence of particles. In addition to the enhanced VLSMs observed in the outer layer, the large-scale energetic structures (e.g. the ``bootleg" within $y^+<100$ in figure \ref{fig:Euu_3d_550}(b,e)) extend into the inner layer, which are referred to as deep $u-$modes \citep{del2003spectra} or VLSM ``footprints" \citep{hutchins2007evidence}, and are a possible path for the inverse scale transfer from LSMs to VLSMs found by \cite{lee2019spectral}.
\\

\begin{figure}
\centering
\putfig{}{\includegraphics[width=13.5cm,trim={0cm 0cm 0cm 0}, clip]{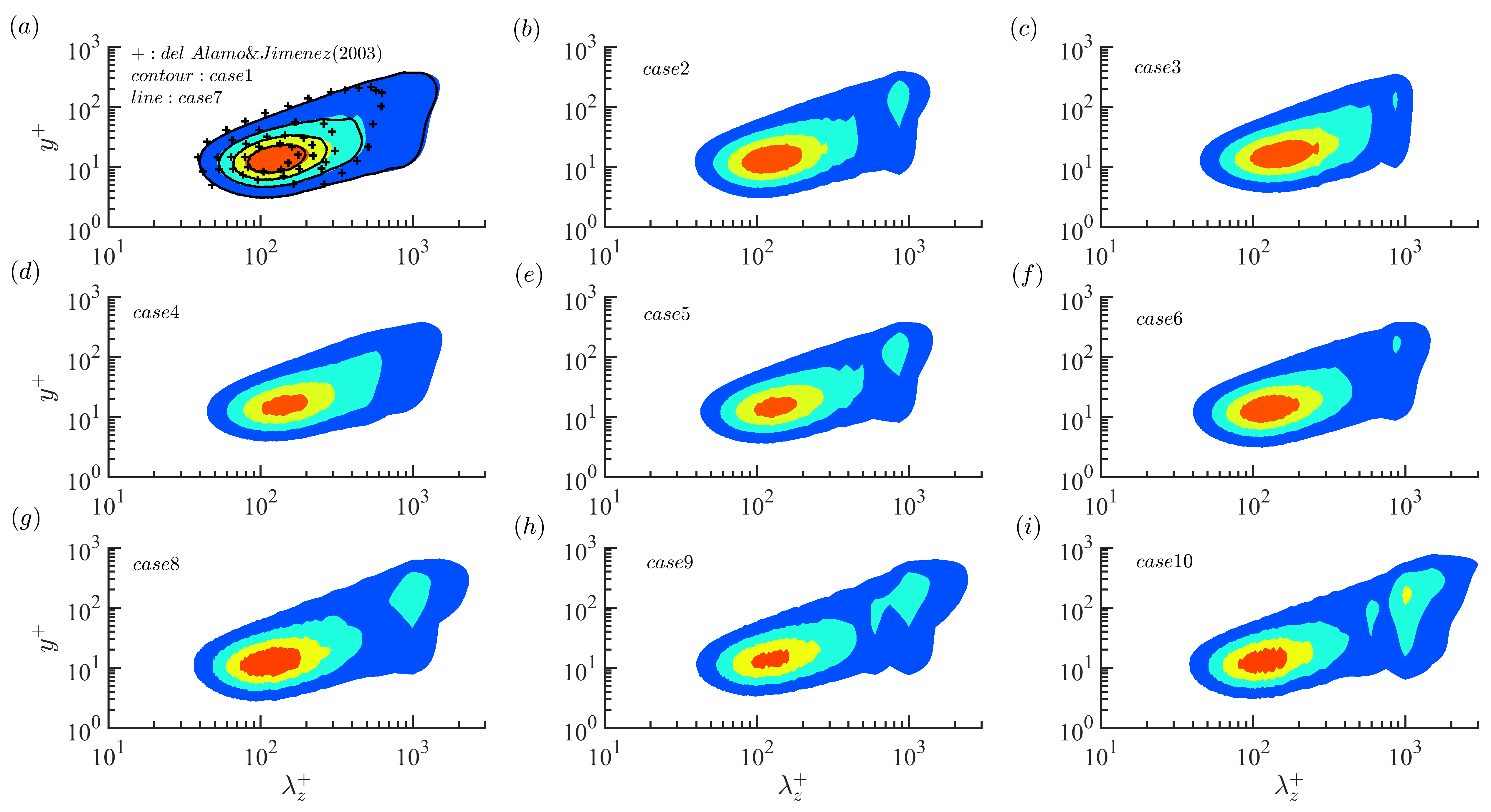}}
  \caption{Premultiplied $1D$ $u$-spectra as functions of spanwise wavelength and wall-normal direction. The contours are $0.2,0.4,0.8$ times the maximum value based on the single-phase flow. $(a-f)$: cases $1-6$ for $\Rey_\tau=550$; $(g-i)$: cases $8-10$ at $\Rey_\tau=950$. In $(a)$: Shaded contours are for $case1$ and line types represent $case7$ which doubles the domain size of $case1$ in streamwise direction. $``+"$: isolines of ($0.2,0.4,0.8$) from \cite{del2003spectra}, containing modes only with $\lambda_x<5h$ (VLSMs with $\lambda_x>5h$ are artificially removed).}
\label{fig:kzEuu_2d_550_950}
\end{figure}

Figure \ref{fig:kzEuu_2d_550_950} displays as a function of $y^{+}$ the premultiplied, one-dimensional $u-$spectra $k_z \phi_{u'u'}(k_z)$, where $\phi_{u'u'}(k_z)=\langle{\hat{u'}(k_z)\hat{u'}^*(k_z)}\rangle$, as a function of the normalized spanwise wavelength $\lambda_{z}^{+}$. As a reference, we compare with the results of \cite{del2003spectra} for wall-bounded channel flow  at the same $\Rey_\uptau=550$, who find that the turbulence in the outer flow behaves roughly isotropic if VLSMs are artificially removed (i.e. VLSMs introduce anisotropy). 
Comparing between \cite{del2003spectra} (who filter out high wavelengths, so $k_z \phi_{u'u'}(k_z)$ with only $\lambda_x<5h$ are plotted) and the present (unfiltered) simulation in figures \ref{fig:kzEuu_2d_550_950}(a), we observe that for unladen flow, with or without the contribution from turbulent structures of $\lambda_x > 5h$, the spectral signature of LSMs is hardly affected (see the spectrum below $y^+=100$ with $\lambda^+_z<200$) \citep{jimenez1999autonomous}. As noted previously, any effect of a limited streamwise domain extent $L_{x}$ is minimal, since figure \ref{fig:kzEuu_2d_550_950}(a) shows that the energy contained in VLSMs in a short domain (contour represents for $case1$) is nearly identical to the long domain ($case7$ is shown by lines). Thus overall our unladen simulations are consistent with the current understanding of VLSMs. We point out that in single-phase flow, only $(0.2,0.4,0.8)$ times the maximum value based on the single-phase flow are displayed in figure \ref{fig:kzEuu_2d_550_950}(a); therefore the second peak in the outer layer is not as readily observed as in figure \ref{fig:Euu_3d_550}(a). For $case2$ (figure \ref{fig:kzEuu_2d_550_950}(b)) and $case5$ (figure \ref{fig:kzEuu_2d_550_950}(e)), the contribution from VLSMs forms a bimodal spanwise spectra at $600<\lambda_z^+ <1000$ ($h< \lambda_z < 2h$) at heights above the inner layer. The enhancement of the VLSM signature is also found at $\Rey_\uptau=950$, $St_{out}=8.0$ as shown in figures \ref{fig:kzEuu_2d_550_950}(g-i) for $cases~8-10$. The Stokes number based on the outer time scale for $case9$ is similar to that of $case5$ at $\Rey_{\uptau}=550$. By investigating inertial particle modulation of the regeneration cycle of LSMs, \cite{wang2019modulation} found that particle inertia has a non-monotonic effect on LSMs in the inner region: low inertia ($St^+=\mathcal{O}(1)$, e.g. $St^+=2.42$ of $case2$) promote the regeneration cycle whereas high inertia ($St^+=\mathcal{O}(10)$, i.e. $St^+=24.2-182$ of $cases~3-5$) attenuate the regeneration cycle. However we see from figure \ref{fig:kzEuu_2d_550_950} that VLSM enhancement occurs at both low and high Stoke numbers (i.e. $case2$ and $case5$). \\

\begin{figure}
\centering
\putfig{}{\includegraphics[width=13.5cm,trim={0cm 0cm 0cm 0cm}, clip]{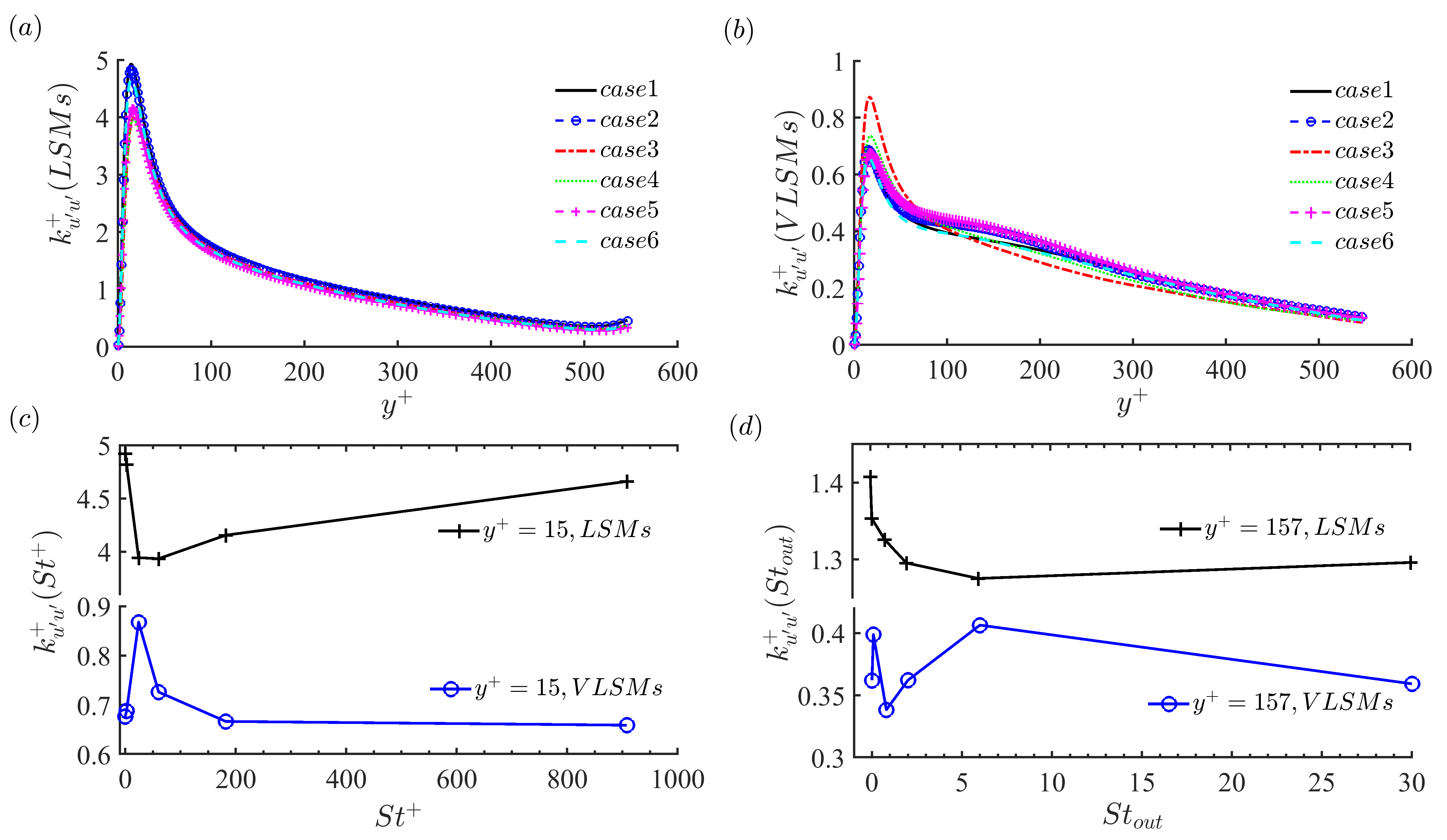}}
  \caption{Streamwise TKE ($k_{u'u'}$) contributed by $(a)$ LSMs, $\lambda_x < 5h $ and $(b)$ VLSMs, $\lambda_x > 5h $ of $case1-6$. $y^+=15$ and $y^+=157$ are used to represent the inner layer and outer layer, respectively. The corresponding value of $k_{u'u'}$ at two heights is plotted against $(c)$ $St^+$ and $(d)$ $St_{out}$.}
\label{fig:Euu_VLSM_LSM_St}
\end{figure}

\begin{figure}
\centering
\putfig{}{\includegraphics[width=13.5cm,trim={0cm 0cm 0cm 0cm}, clip]{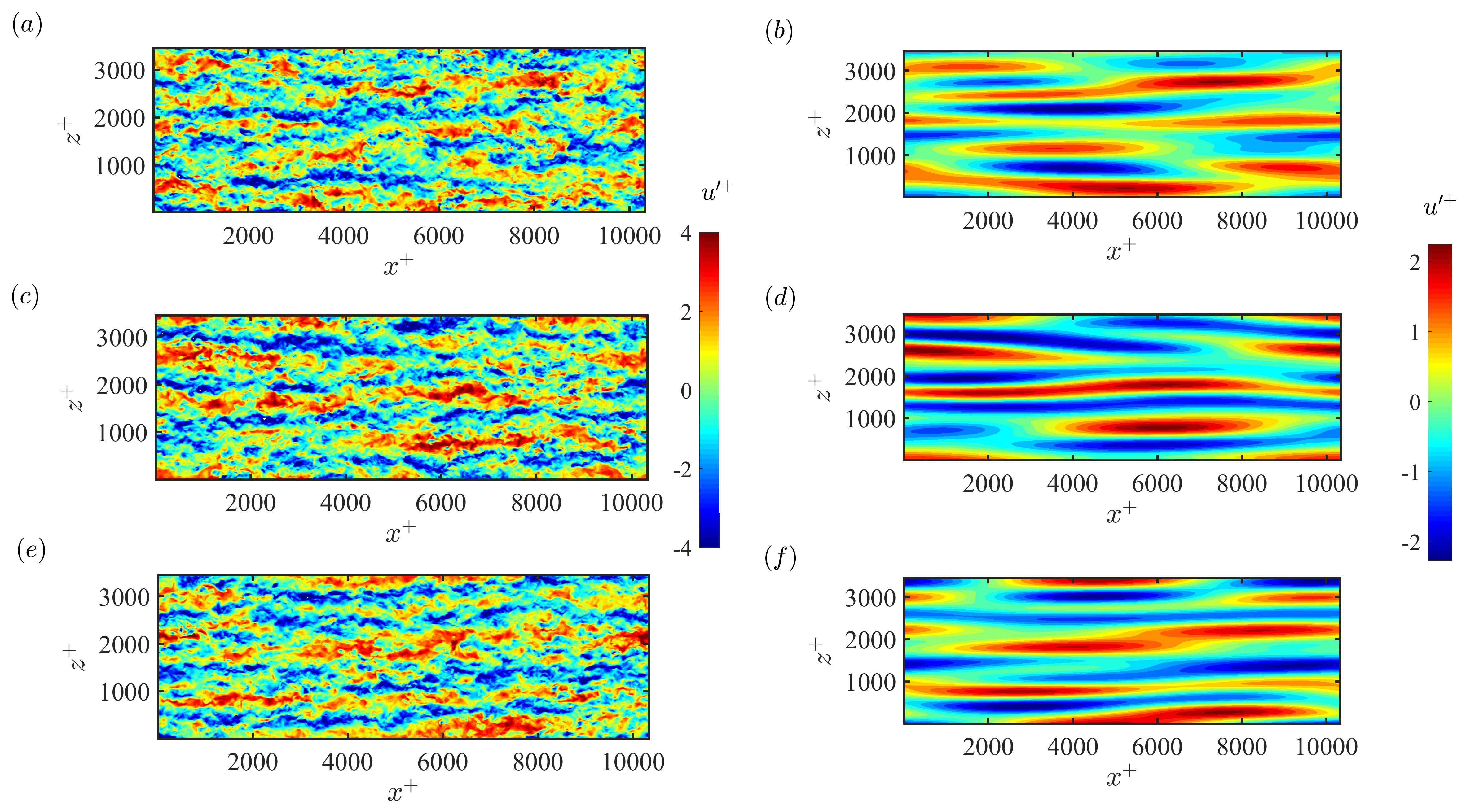}}
  \caption{Instantaneous contours of streamwise velocity fluctuation scaled by $u_\tau$ on a wall-parallel plane at $y^+=157$ at the same time step. $(a,c,e)$ containing all modes. $(b,d,f)$ the same flow field but only associating with VLSMs, containing modes with $\lambda_x > 10h,~\lambda_z > 0.75h$ ($\lambda^+_x > 5500,~\lambda^+_z > 412$). $(a,b)$ $case1$; $(c,d)$ $case2$; $(e,f)$ $case5$.}
\label{fig:uprime_outerflow}
\end{figure}


In order to quantify the TKE of VLSMs and LSMs modulated by particles and demonstrate their effect as function of Stokes number, we show $k_{u'u'} \equiv \overline{u'u'}$ for LSMs (represented by $\lambda_x < 5h$) and VLSMs (represented by $\lambda_x > 5h$) in figures \ref{fig:Euu_VLSM_LSM_St}(a) and (b), respectively. The TKE of LSMs is $4-5$ times larger than the TKE of VLSMs. The $k_{u'u'}$ contributed by LSMs and VLSMs is shown as a function of Stokes number for the representative inner layer height $y^+=15$ in figure \ref{fig:Euu_VLSM_LSM_St}(c) and the representative outer layer height $y^+=157$ in figure \ref{fig:Euu_VLSM_LSM_St}(d). Figure \ref{fig:Euu_VLSM_LSM_St}(c) shows that in the inner layer, the minimum TKE of LSMs appears for $case3$ ($St^+=24.2$), corresponding to the strongest turbulence attenuation observed by \cite{wang2019modulation}. In the outer layer, the minimum TKE of LSMs appears at a higher Stokes number, i.e. $case5$ ($St_{out}=6.0$). However, the TKE modulation of VLSMs is distinct, and even opposite in behaviour, from that of LSMs. As shown in figure \ref{fig:Euu_VLSM_LSM_St}(c), in the inner layer, the intensity of VLSMs reaches its maximum for $case3$ ($St^+=24.2$) whereas the TKE of VLSMs has two peaks in the outer layer: $case2$ ($St_{out}=0.08$) and $case5$ ($St_{out}=6.0$). As argued below, these two peaks correspond to indirect and direct modulation mechanisms, respectively, and will be discussed in section \ref{subsec:Two mechanisms}. \\


Additional qualitative evidence of VLSM enhancement can be seen in figure \ref{fig:uprime_outerflow}, which provides a representative snapshot of the streamwise velocity fluctuation in the $x-z$ plane at $y^+=157$. Comparing the snapshots in figures \ref{fig:uprime_outerflow}(c,e) and (a), it may not easy to detect VLSM modulation. However when the flow field is filtered by a threshold $\lambda_x > 10h$ and $\lambda_z > 0.75h$, as displayed in figures \ref{fig:uprime_outerflow}(b,d,f), it is evident that VLSMs are stronger and more coherent in $case2$ (figure \ref{fig:uprime_outerflow}(d)) and $case5$ (figure \ref{fig:uprime_outerflow}(f)) as compared to single-phase flow. This is a more qualitative confirmation of the results shown in figures \ref{fig:Euu_3d_550}, \ref{fig:kzEuu_2d_550_950}, and \ref{fig:Euu_VLSM_LSM_St}. We now turn our investigation to understanding why the VLSM modulation appears to have two distinct peaks in Stokes number -- in particular demonstrated by $case2$ and $case5$. 

\subsection{Two mechanisms of VLSMs enhancement by particles}\label{subsec:Two mechanisms}

\begin{figure}
\centering
\putfig{}{\includegraphics[width=11cm,trim={1cm 0.5cm 2.5cm 0.0cm}, clip]{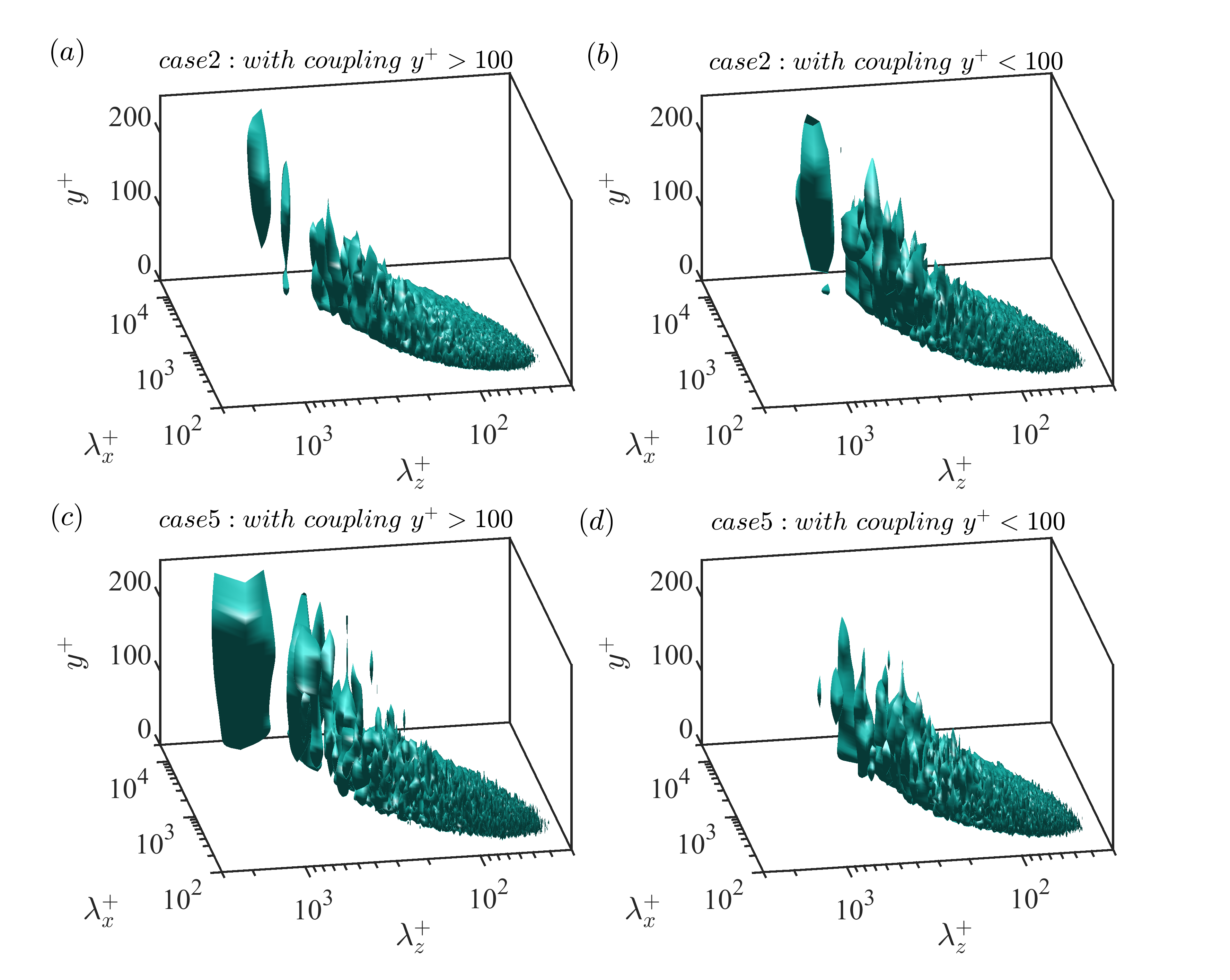}}
  \caption{Premultiplied two-dimensional spectrum $k_xk_z\Phi_{u'u'}/u_\uptau ^2$ as function of $\lambda_x$ and $\lambda_z$ in the wall-normal direction $y$. Isosurface of $0.1$ times the common maximum value based on the single-phase flow is illustrated. $(a,c)$ Only with particle coupling in the outer flow for $case2$ and $case5$; $(b,d)$ Only with particle coupling in the rengeneration cycle region for $case2$ and $case5$.}
\label{fig:test_inner_outer}
\end{figure}

In order to verify the above hypothesis that the VLSM enhancement in $case2$ is due to particles' modulation of LSMs in the inner flow (we refer to this as indirect modulation of VLSMs) whereas in $case5$ it is due to the particles' direct modulation on the VLSMs in the outer flow, we perform a conditional numerical test to identify the particles' effective region of influence regarding VLSM enhancement by artificially applying the particle feedback force only in one of three locations: (1) the viscous sublayer ($y^+<15$), (2) the regeneration cycle region ($15<y^+<100$), or (3) the outer flow ($y^+>100$), separately. The premultiplied two-dimensional energy spectrum of streamwise velocity $k_xk_z\Phi_{u'u'}$ is shown in figure \ref{fig:test_inner_outer} as a function of $y^{+}$. Compared with figure \ref{fig:Euu_3d_550}(a), it can be seen that the VLSMs only experience enhancement in $case2$ when particle coupling is included in the regeneration cycle region. For $case5$, on the other hand, the opposite is true: VLSM enhancement is found only when particle coupling effects are included in the outer region. Both of these effects are observed throughout the entire range of $y^{+}$. The tests of two-way coupling applied for $y^+<15$ are not shown, but we find the spectrum in the range $3.77h < \lambda _x  <  6.28h$ ($2073 <\lambda_x^+ < 3454$) and $ 0.78h < \lambda _z <1.0h$ ($429 <\lambda_z^+ < 550$) is stronger than in the single-phase flow, but shorter and narrower than the streamwise and spanwise scale of the second peak of the TKE spectrum.\\

\begin{figure}
\centering
\putfig{}{\includegraphics[width=13.0cm,trim={0cm 0cm 0cm 0cm}, clip]{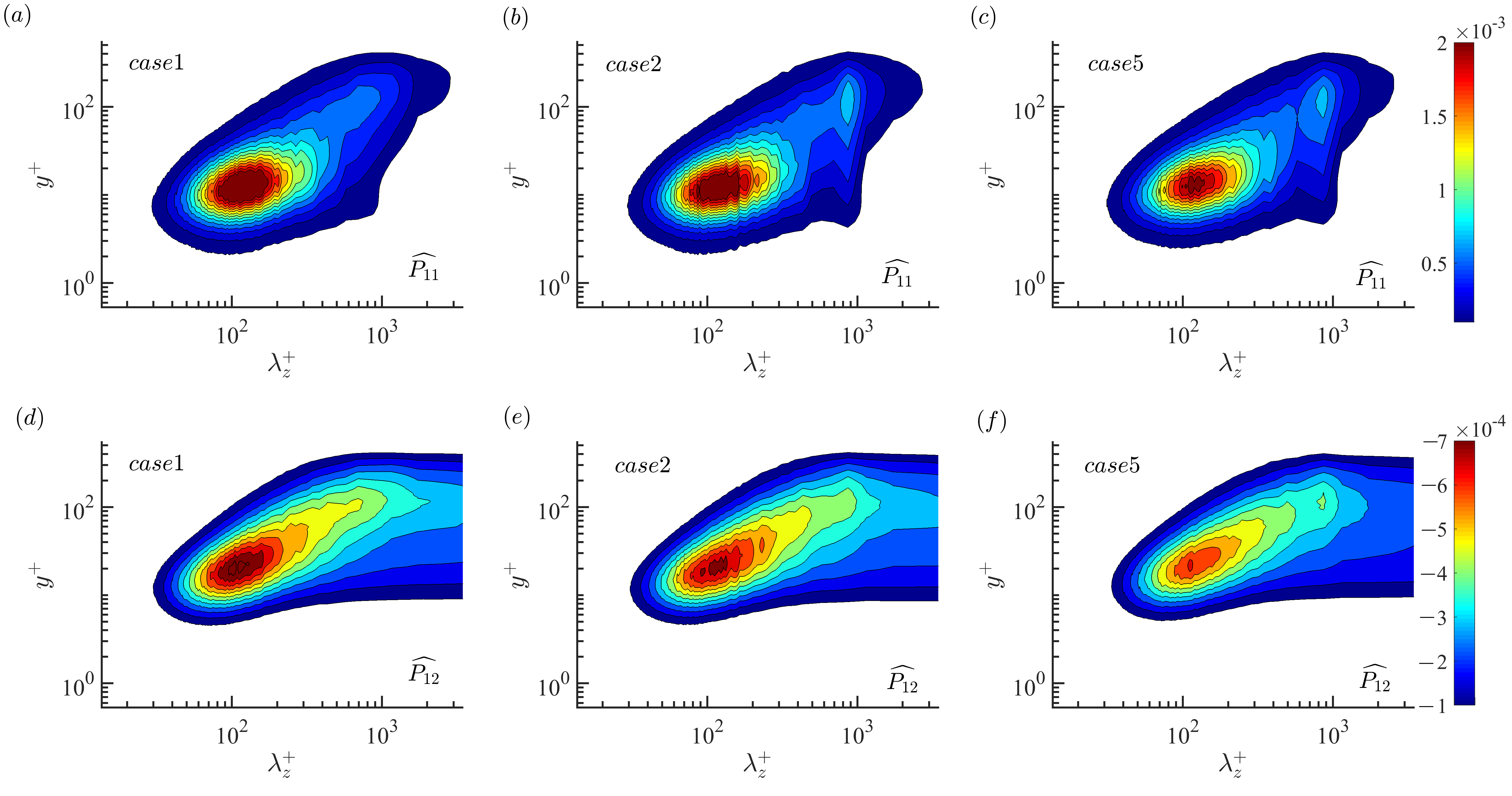}}
  \caption{Production contribution to the streamwise TKE budget and Reynolds shear stress budget in spectral space normalized by $u^3_\uptau/\delta_\nu$. $(a-c)$ $\hat{P}_{11}$ and $(d-f)$ $\hat{P}_{12}$. $(a,d)$ $case1$; $(b,e)$ $case2$; $(c,f)$ $case5$.}
\label{fig:P11_P12}
\end{figure}

In spectral space, the modulation of LSMs and VLSMs by a dispersed phase is at least partially related to the direct influence on velocity fluctuations, which in turn can modify the production of TKE and/or Reynolds shear stress.
This is demonstrated in figure \ref{fig:P11_P12}, where we present the $\overline{u'u'}$ spectral production term $\hat{P}_{11}=-\langle{\hat{u'}(k_z,y)\hat{v'}^*(k_z,y)}\rangle dU/d{y}$ as well as the $\overline{u'v'}$ spectral production term $\hat{P}_{12}=-\langle{\hat{v'}(k_z,y)\hat{v'}^*(k_z,y)}\rangle dU/d{y}$ as a function of $\lambda^+_z$ and the wall-normal direction $y^+$ for two different Stokes numbers both previously seen to enhance VLSMs: low Stokes number $case2$ and high Stokes number $case5$. These are shown in comparison with single-phase flow $case1$. Throughout the wall-normal direction, $\hat{P}_{11}$ is positive whereas $\hat{P}_{12}$ is negative. Comparing figures \ref{fig:P11_P12}(a,b,c) with figures \ref{fig:kzEuu_2d_550_950}(a,b,e) respectively, we find that they have a similar overall shape, and the bimodal spectrum appears both in the premultipled $u-$spectra as well as the production term $\hat{P}_{11}$, which is enhanced in $case2$ and $case5$ in comparison with the single-phase flow. In regards to $\hat{P}_{12}$, a similar overall shape as compared to premultipled $uv$-cospectra (figure is not shown) is observed even though the bimodal spectrum is not as obviously established as that for $\hat{P}_{11}$. The intensity of $\hat{P}_{12}$ ($\hat{P}_{12}$ is negative in the domain) is weakened in $case2$ and $case5$ in comparison with the single-phase flow by the presence of particles. Thus figure \ref{fig:kzEuu_2d_550_950} indicates that production of $\overline{u'u'}$ is enhanced at the heights and wavenumbers associated with VLSMs, while at the same time, particularly for $case5$, the production of Reynolds shear stress is diminished at the same wavelengths and heights.\\

\begin{figure}
\centering
\putfig{}{\includegraphics[width=13.5cm,trim={0cm 0cm 0cm 0cm}, clip]{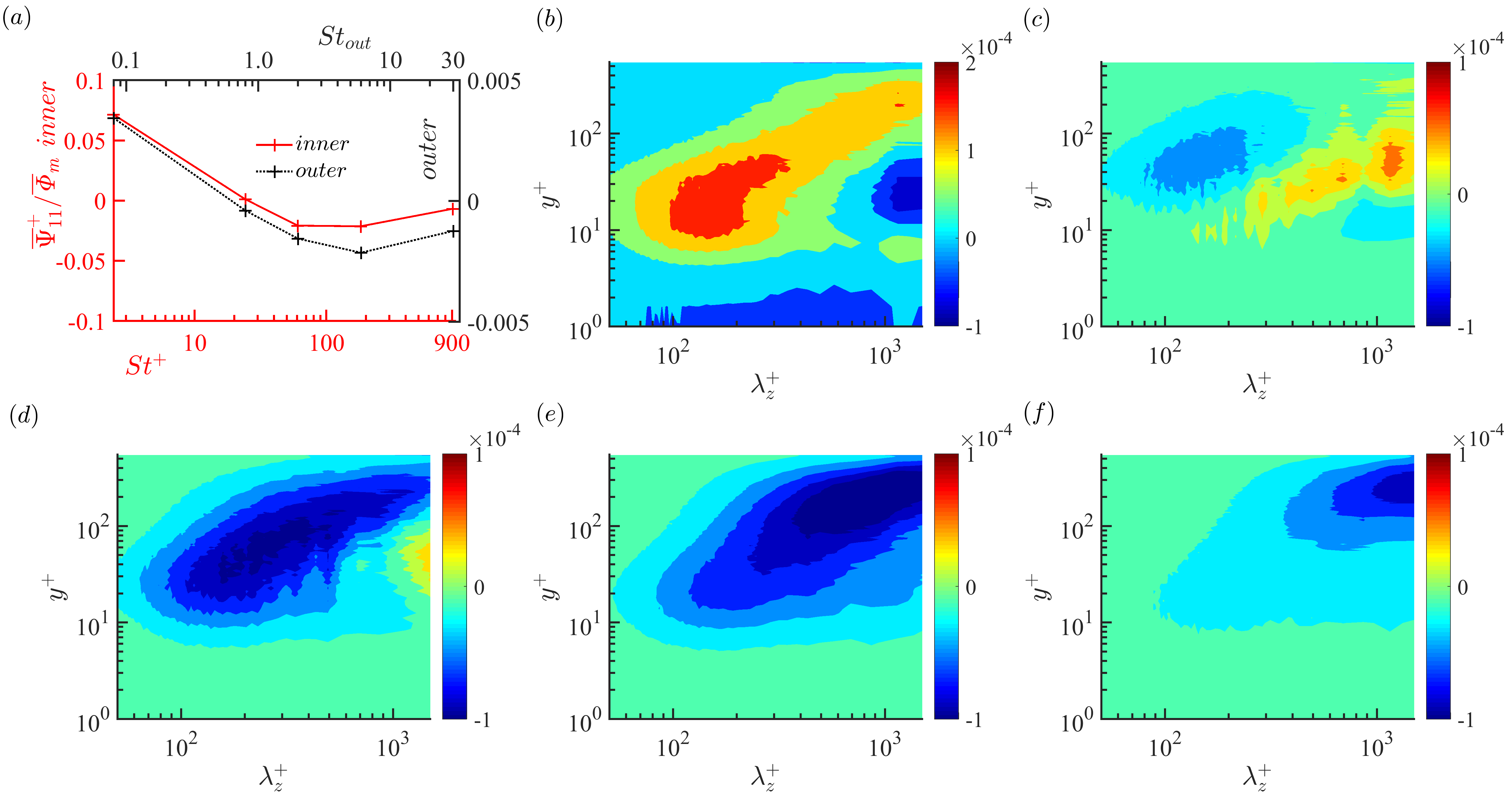}}
  \caption{Particle feedback term ${\Psi}_{11}$ contribution to the streamwise TKE budget. $(a)$ mean value of all modes in the inner and outer layers as a function of Stokes number, normalized by $u^3_\uptau/\delta_\nu$ and bulk mass fraction. $(b-f)$ $\hat{P}_{11}$ contribution in spectral space of cases $2-6$ as functions of spanwise wavelength and wall-normal direction, normalized by $u^3_\uptau/\delta_\nu$ and local mass fraction.}
\label{fig:fft_2d_Fxu}

\centering
\putfig{}{\includegraphics[width=13.5cm,trim={0cm 0cm 0cm 0cm}, clip]{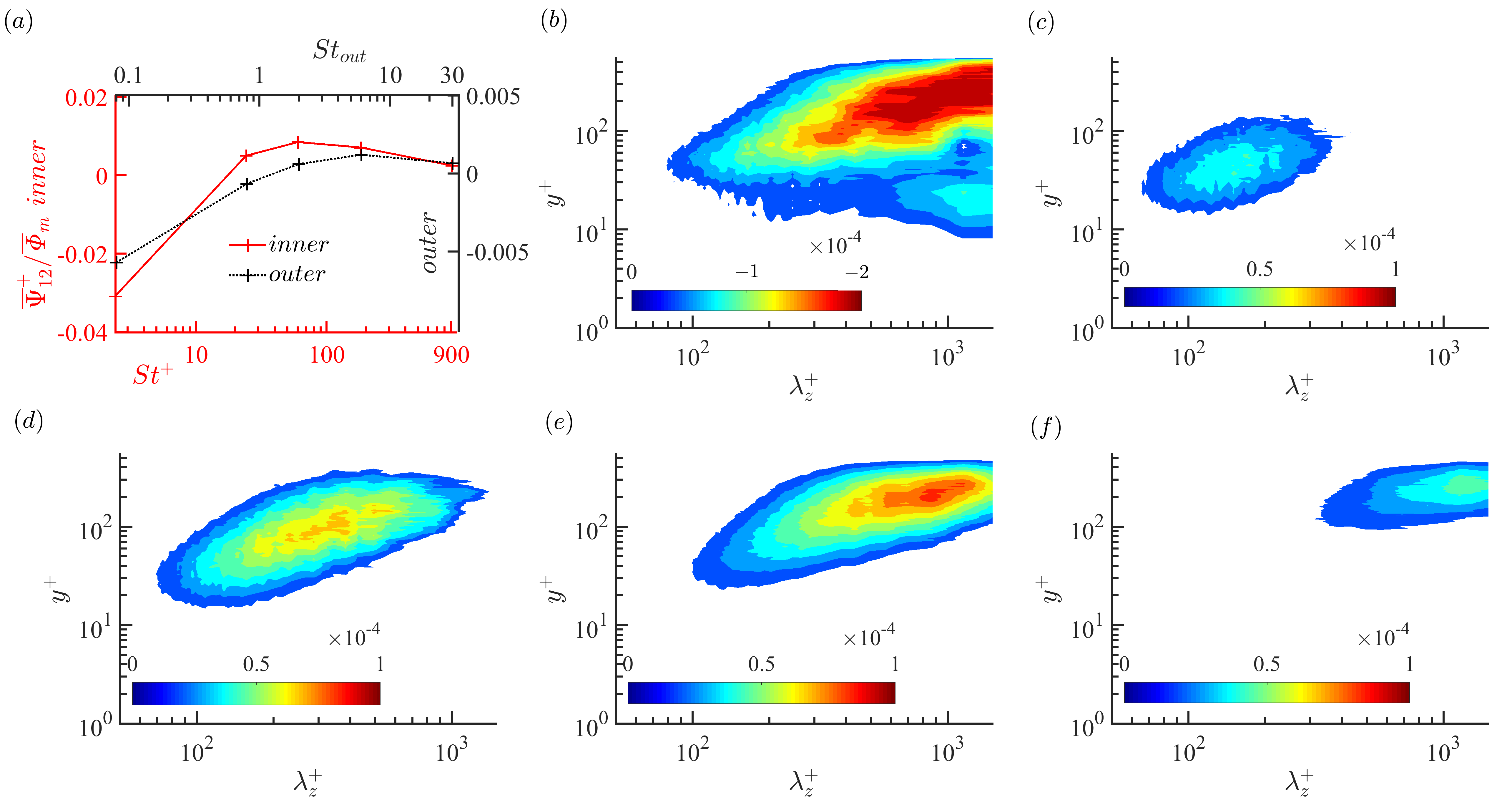}}
  \caption{Particle feedback term ${\Psi}_{12}$ contribution to the Reynolds shear stress budget. Figure legend is the same as figure \ref{fig:fft_2d_Fxu}.}
\label{fig:fft_2d_Fxyuv}
\end{figure}

In addition to the modifications to streamwise TKE and Reynolds shear stress production, particles can also act as a direct source/sink in the spectral TKE and Reynolds stress budgets. In the spectral energy budget, particle sources to the $\overline{u'u'}$ budget are denoted as $\hat{\Psi}_{11}=\mathcal{R} \langle{\hat{F_x'}(k_z,y)\hat{u'}^*(k_z,y)}\rangle$ and to the $\overline{u'v'}$ budget as $\hat{\Psi}_{12}=\mathcal{R} \langle{\hat{F_x'}(k_z,y)\hat{v'}^*(k_z,y)} + {\hat{F_y'}(k_z,y)\hat{u'}^*(k_z,y)}\rangle$, where $\mathcal{R}$ stands for the real part and $\hat{F}$ is the Fourier transform of the particle coupling force.
The mean value of $\Psi_{11} = F'_x u'$ and $\Psi_{12} = F'_x v' + F'_y u'$ (where $F'_i$, $i=x,y,z$ is the fluctuation of the particle feedback force on the carrier phase) in the inner layer and outer layer of all modes normalized by the bulk mass fraction $\Phi_m$, is shown in figure \ref{fig:fft_2d_Fxu}(a) and figure \ref{fig:fft_2d_Fxyuv}(a), respectively. In both the inner layer and outer layer, the sign of ${\Psi}_{11}$ is positive for $case2$ whereas it becomes negative for $cases~3-6$. This is opposite when compared to ${\Psi}_{12}$, indicating that ${\Psi}_{11}$ and ${\Psi}_{12}$ play opposite roles in the streamwise TKE budget and Reynolds stress budgets for the same Stokes number. In the spectral energy budget, the particle source to the $\overline{u'u'}$ budget is denoted by $\hat{\Psi}_{11}$ and to the $\overline{u'v'}$ budget is $\hat{\Psi}_{12}$ (both normalized by local particle mass fraction $\Phi_m$ in figures \ref{fig:fft_2d_Fxu} and \ref{fig:fft_2d_Fxyuv}). With increasing Stokes number as shown in figures \ref{fig:fft_2d_Fxu}(b-f) and figures \ref{fig:fft_2d_Fxyuv}(b-f), the regions of the highest magnitudes of $\hat{\Psi}_{11}$ and $\hat{\Psi}_{12}$ shift from the inner layer to the outer layer, and at the same time from low to high wavelength. In particular, it is found that for $case2$, a region of large, positive $\hat{\Psi}_{11}$ (figure \ref{fig:fft_2d_Fxu}(b)) appears in the inner layer with wavelengths associated with LSMs, whereas in $case5$, a region of large, positive $\hat{\Psi}_{12}$ region instead develops at wavelengths associated with VLSMs (figure \ref{fig:fft_2d_Fxyuv}(e)). In contrast, $case2$ exhibits a negative source of $\hat{\Psi}_{12}$ at these wavelengths (figure \ref{fig:fft_2d_Fxyuv}(b)), but for $case5$ there is at the same time a noticeable change in the sign of the contribution and location of $\hat{\Psi}_{11}$ (figure \ref{fig:fft_2d_Fxu}(e)). This picture is consistent with the conditional tests above that $case2$ works with LSMs in the inner layer whereas $case5$ works with VLSMs in the outer layer. This nearly opposite behavior indicates that there might be two underpinning mechanisms of VLSM enhancement induced by low and high Stokes numbers particles.\\

\begin{figure}
\centering
\putfig{}{\includegraphics[width=13.0cm,trim={0cm 0cm 0cm 0cm}, clip]{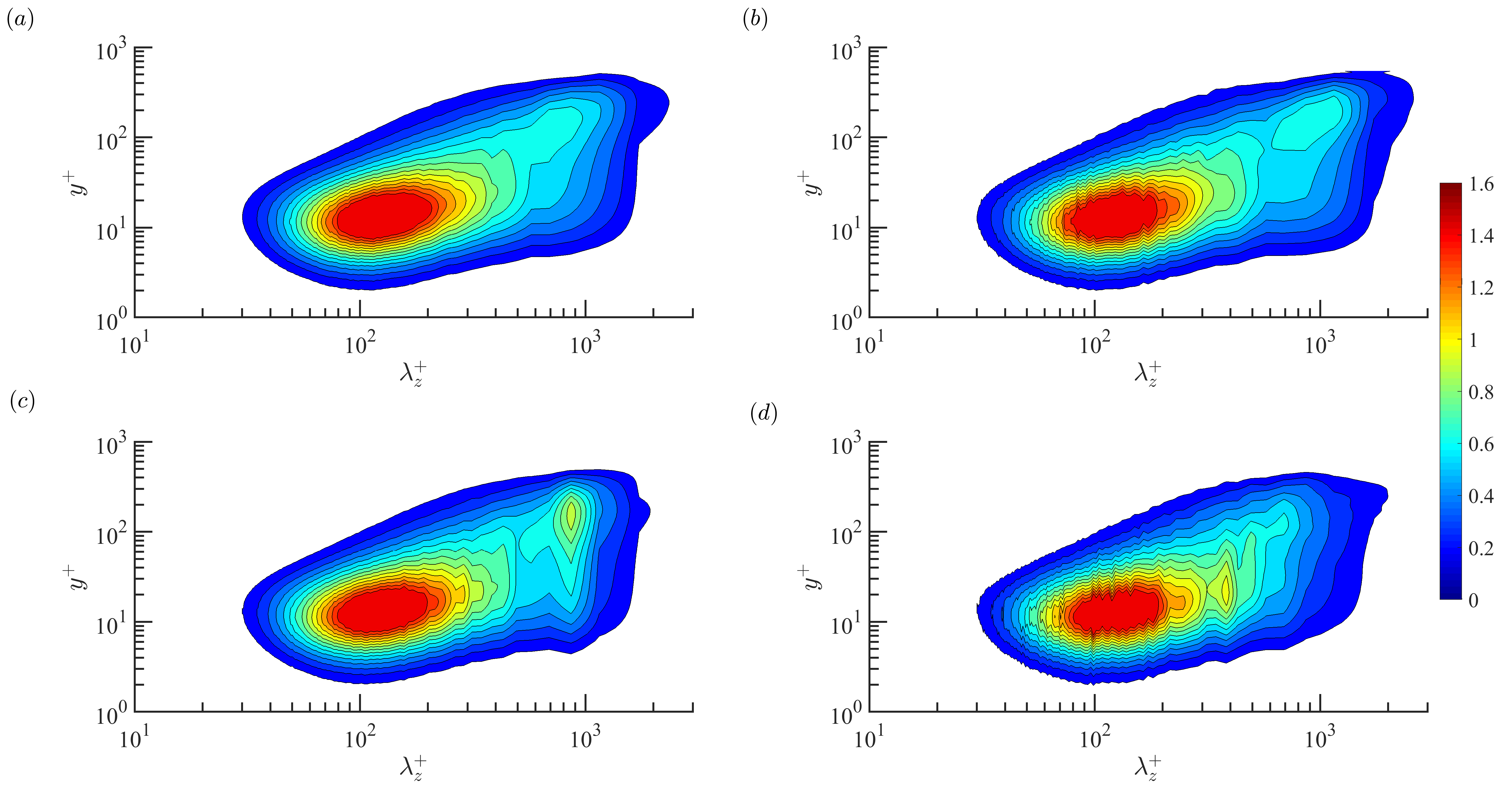}}
  \caption{Premultiplied $1D$ $u$-spectra as functions of spanwise wavelength and wall-normal direction. $(a)$ single phase flow $case1$; $(b-c)$ $St^+=2.42$ of $case2$ with three mass fractions: $(b)$ $\overline{\Phi_m}=1.2 \times 10^{-2}$; $(c)$ $\overline{\Phi_m}=2.4 \times 10^{-2}$; $(d)$ $\overline{\Phi_m}=14 \times 10^{-2}$.}
\label{fig:Euu_kz_diff_Phi}
\end{figure}

In particle-laden flow, \cite{richter2015turbulence} finds that the particle-induced, Stokes-number-dependent source/sink of TKE varies with wavelength, and is possibly associated with the particle clusters themselves \citep{capecelatro2018transition}. Focusing exclusively on the inner region, \cite{wang2019modulation} find that low inertia particles enhance LSMs whereas high inertia particles attenuate LSMs, which can help explain the positive particle feedback $\hat{\Psi}_{11}$ for $case2$ (figure \ref{fig:fft_2d_Fxu}(b)) but negative for $case5$ (figure \ref{fig:fft_2d_Fxu}(e)). In addition, as shown in figure \ref{fig:fft_2d_Fxyuv}(b), the particle source $\hat{\Psi}_{12}$ in $case2$ always attenuates the generation of $\overline{u'v'}$. Therefore in $case2$, the positive feedback $\hat{\Psi}_{11}$ in the inner region is the most likely mechanism responsible for the enhancement of VLSMs in the outer region. This process of particles inducing upscale energy transfer (or a reverse cascade), tending to build up the energy level at high wavelengths due to the modulation of small-scale turbulent motions, is also observed in homogeneous turbulence (see \cite{elghobashi1993two} and \cite{carter2018small}) and in turbulent Couette flow (see \cite{richter2015turbulence}). \\

At low Stokes number ($St^+=\mathcal{O}(1)$), \cite{klinkenberg2013numerical} found that the disturbance energy needed to induce turbulence is low at small mass fractions (i.e. $\overline{\Phi_m} = 0.02-0.06$), whereas is high at large mass fraction (i.e. $\overline{\Phi_m} = 0.138$), indicating a subtle dependence of two-way coupling effects on mass fraction at low Stokes number. In the current context, this understanding can be used to better interpret and understand the indirect modulation of VLSMs by low Stokes number particles. Figure \ref{fig:Euu_kz_diff_Phi} shows one-dimensional $u-$spectra $k_z \phi_{u'u'}(k_z)$ with increasing mass fraction ranging from $\overline{\Phi_m}=1.2 \times 10^{-2}$ to $12 \times 10^{-2}$, all compared with the unladen flow. We observe that the modulation of VLSMs as a function of mass fraction is consistent with the behaviour shown by \cite{klinkenberg2013numerical}: VLSMs are slightly enhanced by a small mass fraction of low Stokes number particles ($\overline{\Phi_m} = 1.2 \times 10^{-2}$ in figure \ref{fig:Euu_kz_diff_Phi}(b)) and significantly promoted by increasing the mass fraction to $\overline{\Phi_m} = 2.4 \times 10^{-2}$ (figure \ref{fig:Euu_kz_diff_Phi}(c)). With further increases in mass fraction, however ($\overline{\Phi_m} = 12 \times 10^{-2}$ in figure \ref{fig:Euu_kz_diff_Phi}(d)), this enhancement begins to diminish. Again, the non-monotonic response of LSM regeneration found in \citet{klinkenberg2013numerical} and \citet{wang2019modulation} appears to be linked to the underlying non-monotonic response of VLSM enhancement to low Stokes number particles.

Finally, in contrast to $case2$, it is more straightforward to understand the VLSM modulation in $case5$. As shown in figure \ref{fig:fft_2d_Fxyuv}(e), throughout the whole domain we observe that $\hat{\Psi}_{12}$ for $case5$ always has a positive contribution to the $\overline{u'v'}$ budget and is at the same spatial locations as the VLSMs of the $u$-spectra (seen also at $\Rey_\uptau=950$ in $case9$, figure is not shown). On the other hand, $\hat{\Psi}_{11}$ for $case5$ in figure \ref{fig:fft_2d_Fxu}(e) tends to suppress the generation of $\overline{u'u'}$. This ultimately results in $\hat{\Psi}_{12}$ exerted in the outer region as being the most likely explanation for the enhancement of VLSMs. As the source of Reynolds shear stress $-\overline{u'v'}$, the `upwelling' and `downwelling' cycles at very large scales \citep{adrian2012coherent} are directly enhanced by the presence of high inertia particles ($St_{out}=6$ of $case5$ at $\Rey_\uptau=550$ and $St_{out}=8.2$ of $case10$ at $\Rey_\uptau=950$) in the outer flow. These very large upwelling/downwelling structures further extract energy from the mean flow by working with local mean shear, as the production of streamwise turbulent kinetic energy budget (${P}_{11}=-\overline{u'v'} dU/d{y}$); see for example the physical explanation from \cite{nezu2005open}.\\

\section{Conclusions}\label{sec:Discussion}
In this paper, we have studied the effect of inertial particles on VLSMs in moderate Reynolds number in open channel flow. Higher particle concentrations are observed in the inner layer than the outer layer due to the towards-wall particle flux induced by turbophoresis. The particle concentration has a non-monotonic dependence on Stokes number whereas the trend is opposite between the inner layer and outer layer. In the inner layer, the particles are characterized by well-known preferential accumulation patterns in the anisotropic LSMs, especially in low-speed streaks, and this behaviour scales as $St^+$ based on inner units. However with increasing wall-normal distance, additional clustering structures are formed in the outer flow. The clustering behaviour is found to be dependent on $St_K$ based on the local Kolmogorov scale, similar to the traditional picture described in isotropic turbulence. In addition, while particles preferentially accumulate in `upwelling' LSMs within the inner layer (especially at $St^+=24.2$ in $case3$), in the outer layer they cluster both in `upwelling' and `downwelling' VLSMs (especially at $St_{out}=6.0$ in $case5$). The distinct bulk concentration and clustering behavior in the two layers are non-monotonically dependent on Stokes number, thereby influencing two-way coupling. This is observed primarily in spectral analysis, where we observe that inertial particles have a non-monotonic effect on the VLSM modulation: low and high inertia particles both strengthen the VLSMs but the intermediate inertia particles hardly affect their structure and energy.\\

By utilizing a conditional numerical test, we demonstrate there are two distinct routes through which inertial particles enhance the VLSMs in the outer layer: low inertia ($St^+=2.42$ based on the inner scale) particles strengthen the VLSMs due to the enhancement of the LSMs in the inner flow. On the contrary, high inertia ($St_{out}= 6.0,8.2$ based on the outer scale) particles strengthen the VLSMs due to direct interaction in the outer flow. The most direct route of particle modulation of turbulent motions comes from the particle feedback source in the turbulent energy budget. Correspondingly, we find that low inertia particles have a positive $\hat{\Psi}_{11}$ in the inner flow and a negative $\hat{\Psi}_{12}$ in the outer flow, which is opposite to high inertia particles. While the relationship between near-wall LSMs and the outer-scale VLSMs remains a subject of investigation, this suggests that there can exist an upscale transport of energy possible from LSMs to VLSMs. Inspired by previously observed, non-monotonic modulation of turbulence of low inertia particles in the inner layer with varying mass fraction \citep{klinkenberg2013numerical}, we observe a VLSM modulation pattern with respect to mass fraction in the outer layer which coincides with the turbulence instability response in the inner layer. This is consistent with \cite{toh2005interaction}, who show numerically that LSMs and VLSMs interact in a co-supporting cycle, \cite{marusic2010predictive} who observe experimentally the high degree of velocity fluctuation correlation between the outer flow with the low-frequency content of the inner flow, and \cite{lee2019spectral} who describe an inverse scale transfer from LSMs to VLSMs close to the wall. In contrast, high inertia particles modulate the VLSMs directly, indicated by the particle feedback effect on the Reynolds shear stress budget with the same scale of VLSMs at the same spatial locations.\\
\section{Acknowledgement}\label{sec:Acknowledgement}
The authors acknowledge grants G00003613-ArmyW911NF-17-0366 from the U.S. Army Research Office and N00014-16-1-2472 from the Office of Naval Research. Computational resources were provided by the High Performance Computing Modernization Program (HPCMP), and by the ND Center for Research Computing.

\bibliography{JFM_RAPID_VLSM}

\end{document}